\pdfoutput=1

\documentclass[12pt,a4paper]{article}

\usepackage{ifthen} 
\newboolean{pdflatex}
\setboolean{pdflatex}{true} 

\newboolean{articletitles}
\setboolean{articletitles}{true} 

\newboolean{uprightparticles}
\setboolean{uprightparticles}{false} 

\newboolean{inbibliography}
\setboolean{inbibliography}{false} 

\def\paperauthors{LHCb collaboration} 
\def\paperasciititle{Study of the B0-> rho(770)0 K*(892)^0 decay with an amplitude analysis of B0-> (\pi+\pi-)(K+\pi-) decays} 
\def\paperkeywords{{High Energy Physics}, {LHCb},{Flavour physics},{CP violation}} 
\def\papertitle{Study of the $B^0\to \rho(770)^0 K^*(892)^0$ decay with an amplitude analysis of $B^0\to (\pi^+\pi^-) (K^+\pi^-)$ decays} 
\def\papercopyright{\the\year\ CERN for the benefit of the LHCb collaboration} 
\def\paperlicence{CC-BY-4.0 licence}
\def\paperlicenceurl{https://creativecommons.org/licenses/by/4.0/}

\usepackage{longtable} 
\usepackage{rotating}
\usepackage{arydshln} 

\usepackage[top=1in, bottom=1.25in, left=1in, right=1in]{geometry}

\usepackage{microtype}
\usepackage{lineno}  
\usepackage{xspace} 
\usepackage{caption} 

\usepackage{graphicx}  
\usepackage{color}
\usepackage{colortbl}
\graphicspath{{./figs/}} 

\usepackage{amsmath} 
\usepackage{amssymb}
\usepackage{amsfonts}
\usepackage{upgreek} 

\usepackage{bm}  
\usepackage{multirow} 
\usepackage[utf8]{inputenc}
\usepackage{adjustbox} 
\usepackage{booktabs, array}
\usepackage{siunitx}

\newcommand*\patchAmsMathEnvironmentForLineno[1]{%
\expandafter\let\csname old#1\expandafter\endcsname\csname #1\endcsname
\expandafter\let\csname oldend#1\expandafter\endcsname\csname
end#1\endcsname
 \renewenvironment{#1}%
   {\linenomath\csname old#1\endcsname}%
   {\csname oldend#1\endcsname\endlinenomath}%
}
\newcommand*\patchBothAmsMathEnvironmentsForLineno[1]{%
  \patchAmsMathEnvironmentForLineno{#1}%
  \patchAmsMathEnvironmentForLineno{#1*}%
}
\AtBeginDocument{%
\patchBothAmsMathEnvironmentsForLineno{equation}%
\patchBothAmsMathEnvironmentsForLineno{align}%
\patchBothAmsMathEnvironmentsForLineno{flalign}%
\patchBothAmsMathEnvironmentsForLineno{alignat}%
\patchBothAmsMathEnvironmentsForLineno{gather}%
\patchBothAmsMathEnvironmentsForLineno{multline}%
\patchBothAmsMathEnvironmentsForLineno{eqnarray}%
}

\usepackage{hyperxmp}

\usepackage[pdftex,
            pdfauthor={\paperauthors},
            pdftitle={\paperasciititle},
            pdfkeywords={\paperkeywords},
            pdfcopyright={Copyright (C) \papercopyright},
            pdflicenseurl={\paperlicenceurl}]{hyperref}

\usepackage[colorinlistoftodos,textsize=scriptsize]{todonotes}

\usepackage[all]{hypcap} 
\usepackage{morefloats}

\usepackage{xspace} 
\usepackage{upgreek}

\newcommand{\offsetoverline}[2][0.1em]{\kern #1\overline{\kern -#1 #2}}%

\def\lhcb   {\mbox{LHCb}\xspace}

\def\babar  {\mbox{BaBar}\xspace}
\def\belle  {\mbox{Belle}\xspace}

\def\lhc    {\mbox{LHC}\xspace}

\def\MagUp {\mbox{\em Mag\kern -0.05em Up}\xspace}

\ifthenelse{\boolean{uprightparticles}}%
{

 \def\Ppi         {\ensuremath{\uppi}\xspace}                 
                  
 \def\Prho        {\ensuremath{\uprho}\xspace}

 \def\Pomega      {\ensuremath{\upomega}\xspace}                 

 \def\PDelta      {\ensuremath{\Delta}\xspace}                 
 \def\PXi         {\ensuremath{\Xi}\xspace}                 
 \def\PLambda     {\ensuremath{\Lambda}\xspace}                 
 \def\PSigma      {\ensuremath{\Sigma}\xspace}                 
 \def\POmega      {\ensuremath{\Omega}\xspace}                 
 \def\PUpsilon    {\ensuremath{\Upsilon}\xspace}

 \def\PB      {\ensuremath{\mathrm{B}}\xspace}                 
                  
 \def\PD      {\ensuremath{\mathrm{D}}\xspace}

 \def\PK      {\ensuremath{\mathrm{K}}\xspace}

 \def\Pb      {\ensuremath{\mathrm{b}}\xspace}                 
 \def\Pc      {\ensuremath{\mathrm{c}}\xspace}                 
 \def\Pd      {\ensuremath{\mathrm{d}}\xspace}

 \def\Pi      {\ensuremath{\mathrm{i}}\xspace}

 \def\Pp      {\ensuremath{\mathrm{p}}\xspace}

 \def\Ps      {\ensuremath{\mathrm{s}}\xspace}                 
                  
 \def\Pu      {\ensuremath{\mathrm{u}}\xspace}

}
{

 \def\Ppi         {\ensuremath{\pi}\xspace}                 
                  
 \def\Prho        {\ensuremath{\rho}\xspace}

 \def\Pomega      {\ensuremath{\omega}\xspace}                 
 \mathchardef\PDelta="7101
 \mathchardef\PXi="7104
 \mathchardef\PLambda="7103
 \mathchardef\PSigma="7106
 \mathchardef\POmega="710A
 \mathchardef\PUpsilon="7107
                  
 \def\PB      {\ensuremath{B}\xspace}                 
                  
 \def\PD      {\ensuremath{D}\xspace}

 \def\PK      {\ensuremath{K}\xspace}

 \def\Pb      {\ensuremath{b}\xspace}                 
 \def\Pc      {\ensuremath{c}\xspace}                 
 \def\Pd      {\ensuremath{d}\xspace}

 \def\Pi      {\ensuremath{i}\xspace}

 \def\Pp      {\ensuremath{p}\xspace}

 \def\Ps      {\ensuremath{s}\xspace}                 
                  
 \def\Pu      {\ensuremath{u}\xspace}

}

\makeatletter
\ifcase \@ptsize \relax
  \newcommand{\miniscule}{\@setfontsize\miniscule{4}{5}}
\or
  \newcommand{\miniscule}{\@setfontsize\miniscule{5}{6}}
\or
  \newcommand{\miniscule}{\@setfontsize\miniscule{5}{6}}
\fi
\makeatother

\DeclareRobustCommand{\optbar}[1]{\shortstack{{\miniscule (\rule[.5ex]{1.25em}{.18mm})}
  \\ [-.7ex] $#1$}}

\def\uquark    {{\ensuremath{\Pu}}\xspace}
\def\uquarkbar {{\ensuremath{\overline \uquark}}\xspace}
\def\uubar     {{\ensuremath{\uquark\uquarkbar}}\xspace}
\def\dquark    {{\ensuremath{\Pd}}\xspace}
\def\dquarkbar {{\ensuremath{\overline \dquark}}\xspace}
\def\ddbar     {{\ensuremath{\dquark\dquarkbar}}\xspace}
\def\squark    {{\ensuremath{\Ps}}\xspace}

\def\cquark    {{\ensuremath{\Pc}}\xspace}

\def\bquark    {{\ensuremath{\Pb}}\xspace}

\def\pion   {{\ensuremath{\Ppi}}\xspace}
\def\piz    {{\ensuremath{\pion^0}}\xspace}
\def\pip    {{\ensuremath{\pion^+}}\xspace}
\def\pim    {{\ensuremath{\pion^-}}\xspace}

\def\rhomeson {{\ensuremath{\Prho}}\xspace}
\def\rhoz     {{\ensuremath{\rhomeson^0}}\xspace}
\def\rhop     {{\ensuremath{\rhomeson^+}}\xspace}
\def\rhom     {{\ensuremath{\rhomeson^-}}\xspace}

\def\kaon    {{\ensuremath{\PK}}\xspace}
  \def\Kbar    {{\kern 0.2em\overline{\kern -0.2em \PK}{}}\xspace}

\def\KorKbar {\kern 0.18em\optbar{\kern -0.18em K}{}\xspace}
\def\Kz      {{\ensuremath{\kaon^0}}\xspace}

\def\Kp      {{\ensuremath{\kaon^+}}\xspace}
\def\Km      {{\ensuremath{\kaon^-}}\xspace}

\def\Kstarz  {{\ensuremath{\kaon^{*0}}}\xspace}
\def\Kstarzb {{\ensuremath{\Kbar{}^{*0}}}\xspace}
\def\Kstar   {{\ensuremath{\kaon^*}}\xspace}

\def\Kstarp  {{\ensuremath{\kaon^{*+}}}\xspace}

\newcommand{\omegaz}{\ensuremath{\Pomega}\xspace}

  \def\Dbar    {{\kern 0.2em\overline{\kern -0.2em \PD}{}}\xspace}
\def\D       {{\ensuremath{\PD}}\xspace}

\def\DorDbar {\kern 0.18em\optbar{\kern -0.18em D}{}\xspace}
\def\Dz      {{\ensuremath{\D^0}}\xspace}
\def\Dzb     {{\ensuremath{\Dbar{}^0}}\xspace}

\def\Dm      {{\ensuremath{\D^-}}\xspace}

\def\B       {{\ensuremath{\PB}}\xspace}
\def\Bbar    {{\ensuremath{\kern 0.18em\overline{\kern -0.18em \PB}{}}}\xspace}

\def\BorBbar    {\kern 0.18em\optbar{\kern -0.18em B}{}\xspace}

\def\Bu      {{\ensuremath{\B^+}}\xspace}

\def\Bd      {{\ensuremath{\B^0}}\xspace}
\def\Bs      {{\ensuremath{\B^0_\squark}}\xspace}
\def\Bsb     {{\ensuremath{\Bbar{}^0_\squark}}\xspace}

\def\Bdb     {{\ensuremath{\Bbar{}^0}}\xspace}

\def\Y#1S{\ensuremath{\PUpsilon{(#1S)}}\xspace}

\def\proton      {{\ensuremath{\Pp}}\xspace}

\def\Lz          {{\ensuremath{\PLambda}}\xspace}

\def\LorLbar     {\kern 0.18em\optbar{\kern -0.18em \PLambda}{}\xspace}

\def\Lb           {{\ensuremath{\Lz^0_\bquark}}\xspace}

\newcommand{\decay}[2]{\mbox{\ensuremath{#1\!\to #2}}\xspace}         

\def\to                 {\ensuremath{\rightarrow}\xspace}

\def\P                {{\ensuremath{P}}\xspace}
\def\CP                {{\ensuremath{C\!P}}\xspace}

\def\AT#1     {\ensuremath{A_{\mathrm{T}}^{#1}}\xspace}           

\def\C#1      {\ensuremath{\mathcal{C}_{#1}}\xspace}                       
\def\Cp#1     {\ensuremath{\mathcal{C}_{#1}^{'}}\xspace}                    
\def\Ceff#1   {\ensuremath{\mathcal{C}_{#1}^{\mathrm{(eff)}}}\xspace}        
\def\Cpeff#1  {\ensuremath{\mathcal{C}_{#1}^{'\mathrm{(eff)}}}\xspace}       
\def\Ope#1    {\ensuremath{\mathcal{O}_{#1}}\xspace}                       
\def\Opep#1   {\ensuremath{\mathcal{O}_{#1}^{'}}\xspace}                    

\newcommand{\tev}{\ifthenelse{\boolean{inbibliography}}{\ensuremath{~T\kern -0.05em eV}}{\ensuremath{\mathrm{\,Te\kern -0.1em V}}}\xspace}
\newcommand{\gev}{\ensuremath{\mathrm{\,Ge\kern -0.1em V}}\xspace}
\newcommand{\mev}{\ensuremath{\mathrm{\,Me\kern -0.1em V}}\xspace}
\newcommand{\kev}{\ensuremath{\mathrm{\,ke\kern -0.1em V}}\xspace}
\newcommand{\ev}{\ensuremath{\mathrm{\,e\kern -0.1em V}}\xspace}
\newcommand{\mevc}{\ensuremath{{\mathrm{\,Me\kern -0.1em V\!/}c}}\xspace}
\newcommand{\gevc}{\ensuremath{{\mathrm{\,Ge\kern -0.1em V\!/}c}}\xspace}
\newcommand{\mevcc}{\ensuremath{{\mathrm{\,Me\kern -0.1em V\!/}c^2}}\xspace}
\newcommand{\gevcc}{\ensuremath{{\mathrm{\,Ge\kern -0.1em V\!/}c^2}}\xspace}
\newcommand{\gevgevcc}{\ensuremath{{\mathrm{\,Ge\kern -0.1em V^2\!/}c^2}}\xspace} 
\newcommand{\gevgevcccc}{\ensuremath{{\mathrm{\,Ge\kern -0.1em V^2\!/}c^4}}\xspace} 

\def\mum  {\ensuremath{{\,\upmu\mathrm{m}}}\xspace}

\def\invfb   {\ensuremath{\mbox{\,fb}^{-1}}\xspace}

\newcommand{\chisq}{\ensuremath{\chi^2}\xspace}

\newcommand{\chisqip}{\ensuremath{\chi^2_{\text{IP}}}\xspace}

\def\gsim{{~\raise.15em\hbox{$>$}\kern-.85em
          \lower.35em\hbox{$\sim$}~}\xspace}
\def\lsim{{~\raise.15em\hbox{$<$}\kern-.85em
          \lower.35em\hbox{$\sim$}~}\xspace}

\newcommand{\Real}{\ensuremath{\mathcal{R}e}\xspace}
\newcommand{\Imag}{\ensuremath{\mathcal{I}m}\xspace}

\def\sPlot{\mbox{\em sPlot}\xspace}

\def\sqs   {\ensuremath{\protect\sqrt{s}}\xspace}

\def\pt         {\ensuremath{p_{\mathrm{T}}}\xspace}

\def\ptot       {\ensuremath{p}\xspace}

\def\evtgen     {\mbox{\textsc{EvtGen}}\xspace}

\def\geant      {\mbox{\textsc{Geant4}}\xspace}

\def\photos     {\mbox{\textsc{Photos}}\xspace}

\def\pythia     {\mbox{\textsc{Pythia}}\xspace}

\def\tell1  {TELL1\xspace}
\def\ukl1   {UKL1\xspace}

\def\kpic        {\ensuremath{\Kp\pim}\xspace}
\def\kpim        {\ensuremath{\Km\pip}\xspace}
\def\kpin        {\ensuremath{K\pi}\xspace}
\def\pipin        {\ensuremath{\pi\pi}\xspace}
\def\pipic        {\ensuremath{\pip\pim}\xspace}
\newcommand{\pz}{\ensuremath{\phantom{0}}}
\newcommand{\pzz}{\ensuremath{\phantom{00}}}
\newcommand{\psm}{\ensuremath{\phantom{-}}}

\def\brhokst {\decay{\Bd}{\rhoz\Kstarz}}
\def\bomkst{\decay{\Bd}{\omegaz \Kstarz}}
\def\bdrhokst{\decay{\Bd}{\rhomeson(770)^0 \Kstar(892)^0}}

\def\bskstkst{\decay{\Bs}{\Kstarz \Kstarzb}}
\def\bskpikpi{\decay{\Bs}{(\kpic)(\kpim)}}

\newcommand{\cp}{\CP}

\newcommand{\fzfh}{$f_0(500)$\xspace}
\newcommand{\fznh}{$f_0(980)$\xspace}
\newcommand{\fzhm}{$f_0(1370)$\xspace}

\newcommand{\eq}[1]{Eq.~\ref{eq:#1}}

\newcommand{\tab}[1]{Table~\ref{tab:#1}}
\newcommand{\tabs}[2]{Tables~\ref{tab:#1} and \ref{tab:#2}}
\newcommand{\fig}[1]{Fig.~\ref{fig:#1}}

\newcommand{\app}[1]{Appendix~\ref{app:#1}}
\newcommand{\refsec}[1]{Sect.~\ref{sec:#1}}

\usepackage{cite} 
\usepackage{mciteplus}

\begin{document}

\renewcommand{\thefootnote}{\fnsymbol{footnote}}
\setcounter{footnote}{1}


\begin{titlepage}
\pagenumbering{roman}

\vspace*{-1.5cm}
\centerline{\large EUROPEAN ORGANIZATION FOR NUCLEAR RESEARCH (CERN)}
\vspace*{1.5cm}
\noindent
\begin{tabular*}{\linewidth}{lc@{\extracolsep{\fill}}r@{\extracolsep{0pt}}}
\ifthenelse{\boolean{pdflatex}}
{\vspace*{-1.5cm}\mbox{\!\!\!\includegraphics[width=.14\textwidth]{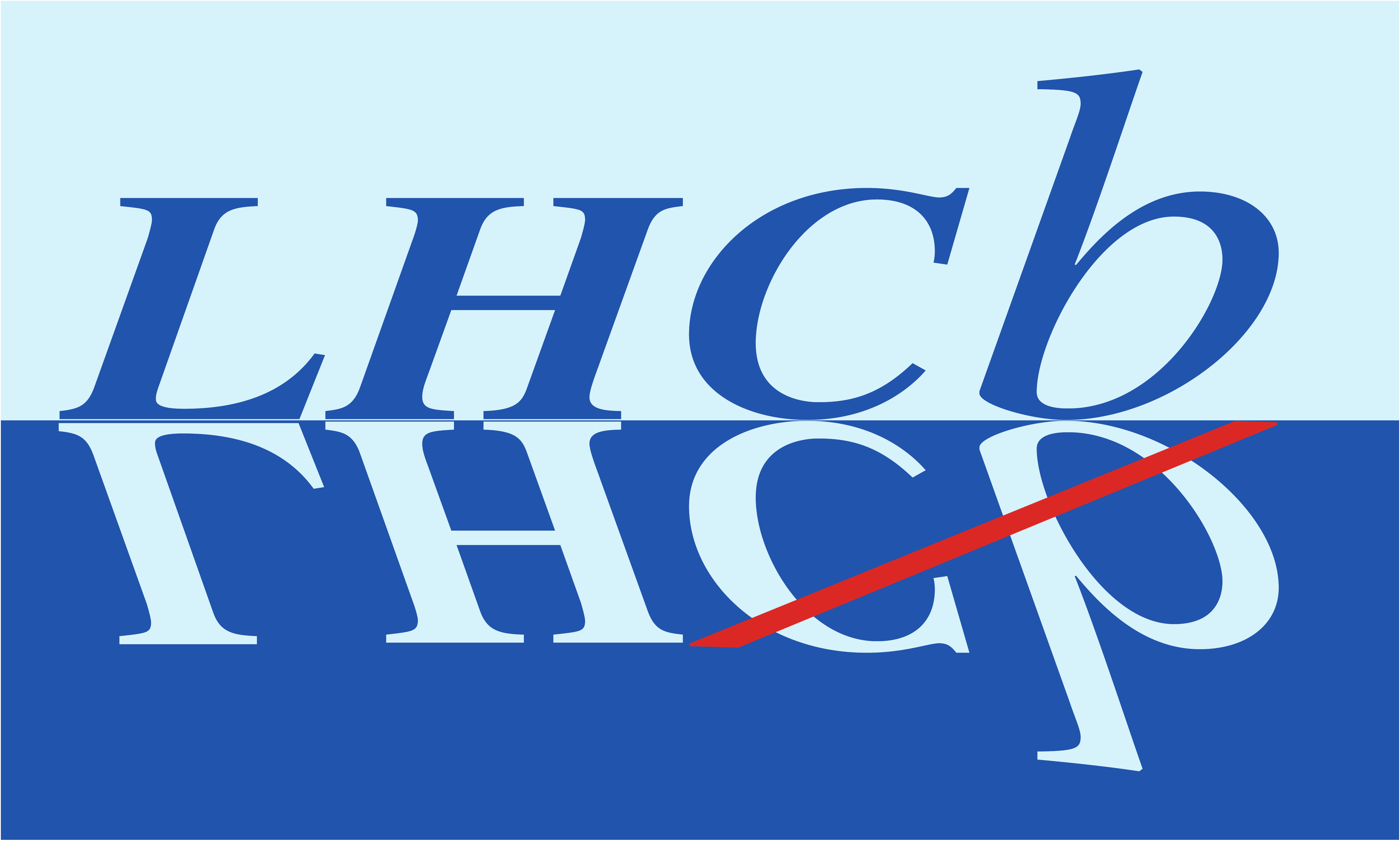}} & &}%
{\vspace*{-1.2cm}\mbox{\!\!\!\includegraphics[width=.12\textwidth]{figs/lhcb-logo.eps}} & &}%
\\
 & & CERN-EP-2018-316 \\  
 & & LHCb-PAPER-2018-042 \\  
 & & 17 December 2018 \\ 
 & & \\
\end{tabular*}

\vspace*{4.0cm}

{\normalfont\bfseries\boldmath\huge
\begin{center}
  \papertitle 
\end{center}
}

\vspace*{2.0cm}

\begin{center}
\paperauthors\footnote{Authors are listed at the end of this paper.}
\end{center}

\vspace{\fill}

\begin{abstract}
  \noindent
 An amplitude analysis of ${B^0\to (\pi^+\pi^-) (K^+\pi^-)}$ decays is performed in the two-body invariant mass regions ${300 < m(\pi^+\pi^-)<1100}$\mevcc, accounting for the $\rho^0$, $\omega$, $f_0(500)$, $f_0(980)$ and $f_0(1370)$ resonances, and ${750 < m(K^+\pi^-)<1200}$\mevcc, which is dominated by the $K^{*}(892)^0$ meson. The analysis uses 3\invfb of proton-proton collision data collected by the \lhcb experiment at centre-of-mass energies of 7 and 8\tev. 
 The \CP averages and asymmetries are measured for the magnitudes and phase differences of the contributing amplitudes. The \ensuremath{C\!P}-averaged longitudinal polarisation fractions of the vector-vector modes are found to be \mbox{$\tilde{f}^0_{\rho K^{*}} = 0.164 \pm 0.015 \pm 0.022$}
and 
\mbox{$\tilde{f}^0_{\omega K^{*}} = 0.68 \pm 0.17 \pm 0.16$}, and their \CP asymmetries, 
\mbox{$\mathcal{A}^0_{\rho K^{*}} = -0.62 \pm 0.09 \pm 0.09$}
and 
\mbox{$\mathcal{A}^0_{\omega K^{*}} = -0.13 \pm 0.27 \pm 0.13$}, where the first uncertainty is statistical and the second systematic.

\end{abstract}

\vspace*{2.0cm}

\begin{center}
Published in JHEP 05 (2019) 026
\end{center}

\vspace{\fill}

{\footnotesize 
\centerline{\copyright~\papercopyright. \href{\paperlicenceurl}{\paperlicence}.}}
\vspace*{2mm}

\end{titlepage}


\newpage
\setcounter{page}{2}
\mbox{~}
%
%
%
%

\cleardoublepage


\renewcommand{\thefootnote}{\arabic{footnote}}
\setcounter{footnote}{0}



\pagestyle{plain} 
\setcounter{page}{1}
\pagenumbering{arabic}


%

\section{Introduction}
\label{sec:intro}
Differences in the behaviour of matter and antimatter (\CP violation) have been observed in several processes and, in particular, in charmless \B decays. The current understanding of the composition of matter in the Universe indicates that other mechanisms, beyond those proposed within the Standard Model (SM) of particle physics, should exist in order to account for the observed imbalance in the matter and antimatter abundances. The study of \CP-violating processes may therefore be used to test the corresponding SM predictions and place constraints on extensions of this framework. In this work, a set of \CP-violating observables is measured using \Bd meson decays reconstructed in the (\pipic)(\kpic) quasi-two-body final state.\footnote{The inclusion of charge conjugate processes is implied.} Particular emphasis is placed on the $\Bd\rightarrow \rho(770)^0\Kstar(892)^0$ decay (hereafter, denoted by \brhokst).

\begin{figure}[t]
\begin{centering}
\includegraphics[width=0.327\textwidth]{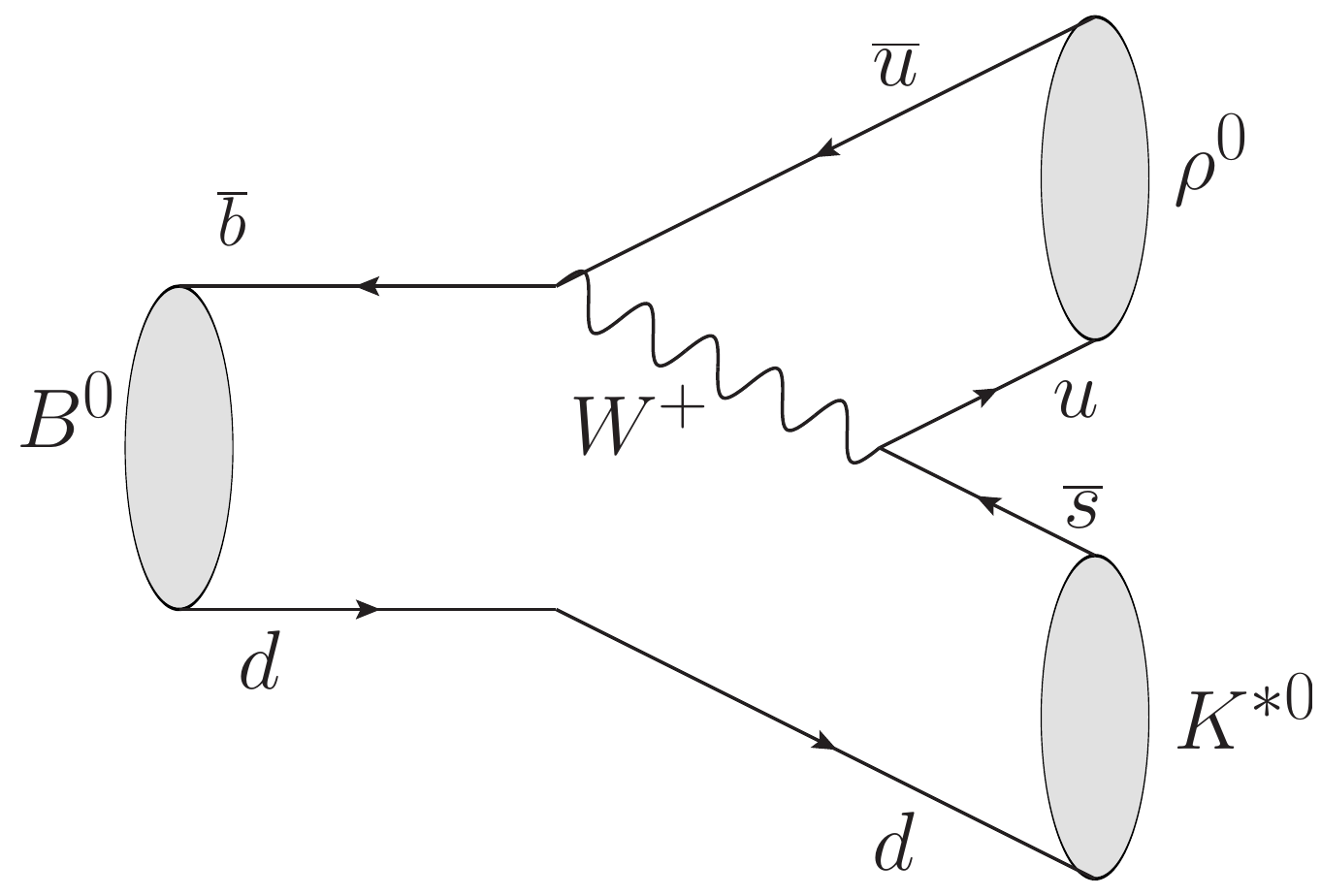}
\includegraphics[width=0.327\textwidth]{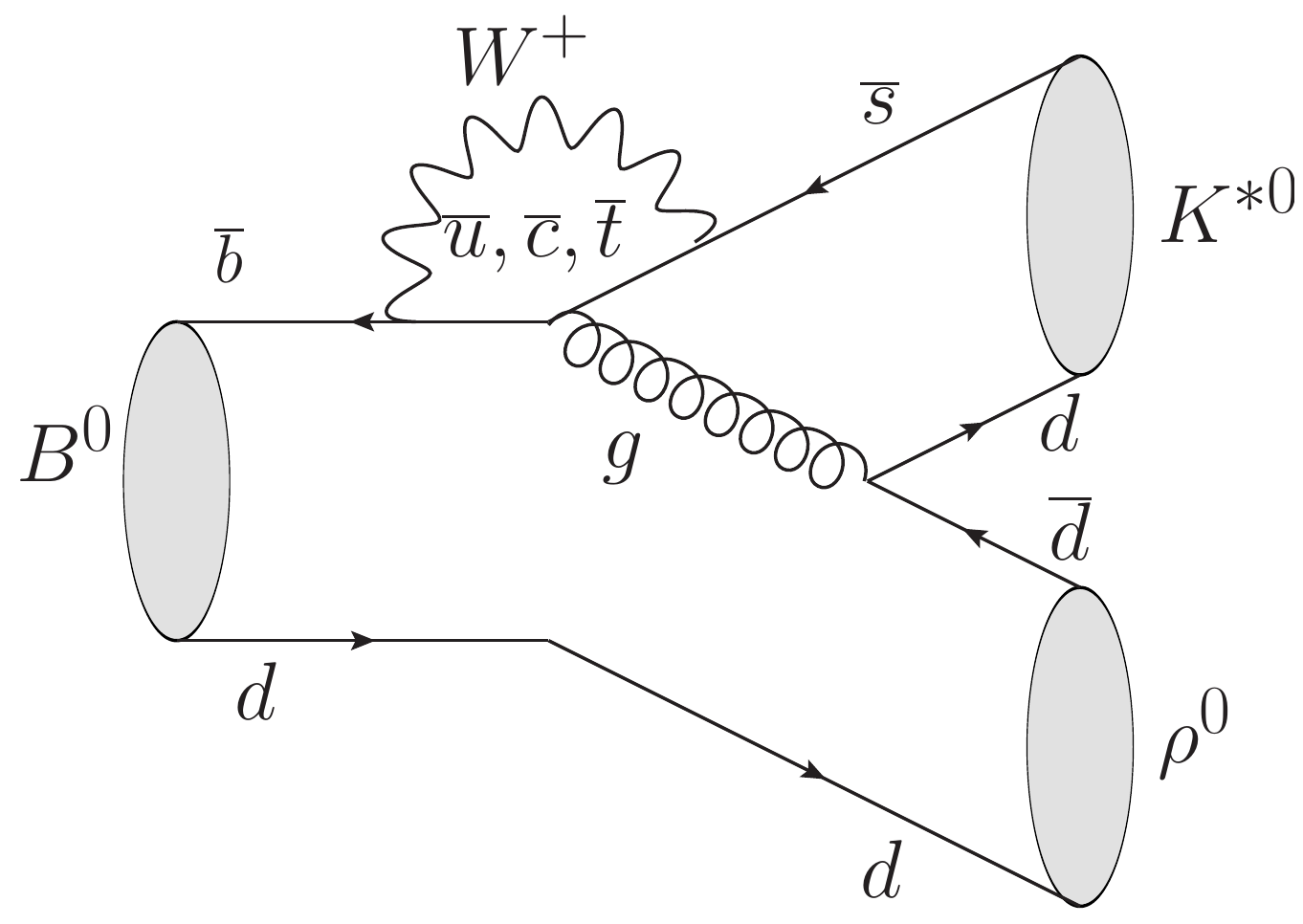}
\includegraphics[width=0.327\textwidth]{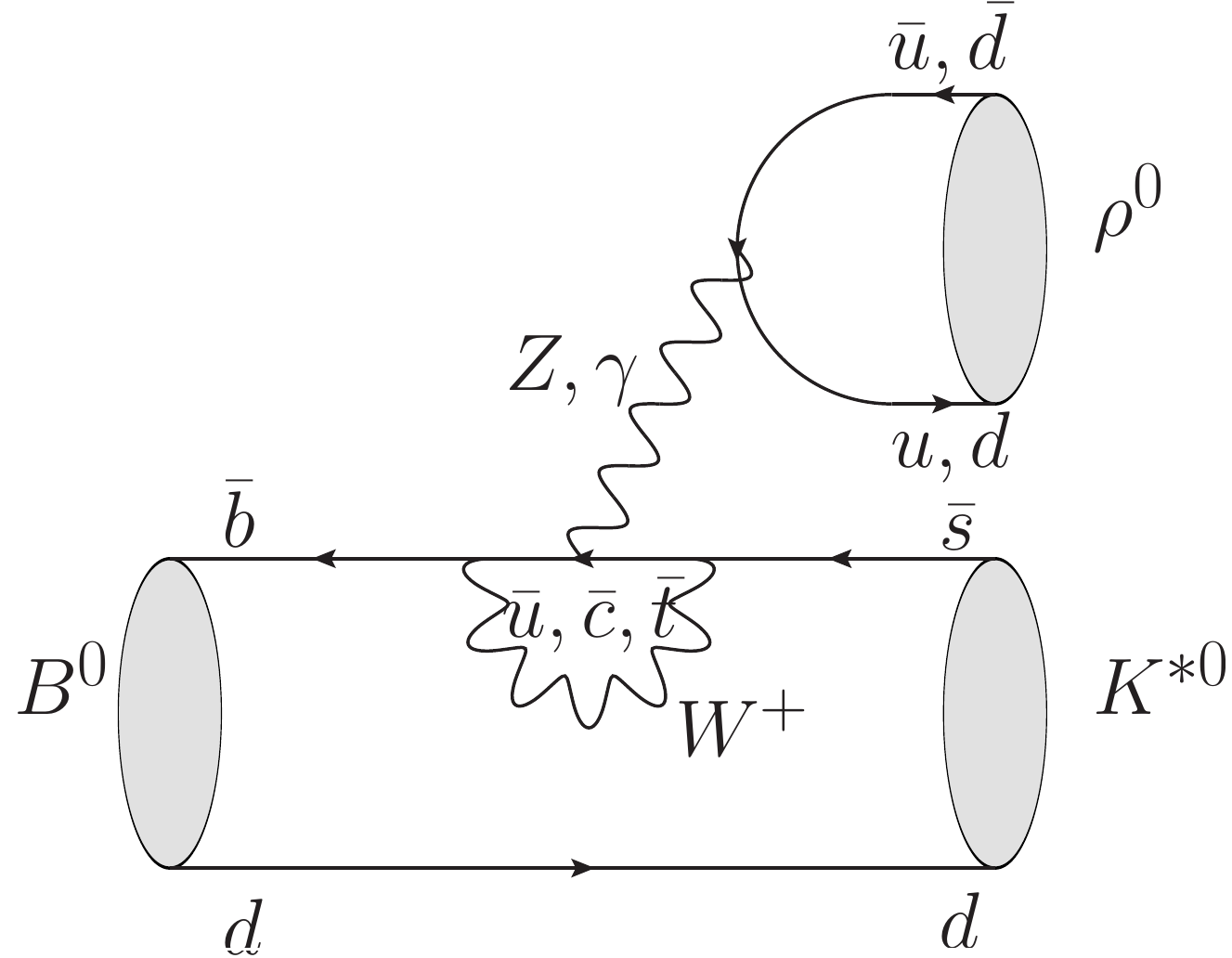}
\caption{\small Leading Feynman diagrams in the \brhokst decay, from left to right: doubly Cabibbo-suppressed tree, gluonic-penguin and electroweak-penguin diagrams.}
\label{fig:diagrams}
\end{centering}
\end{figure}

Direct \CP violation manifests through the difference between partial widths of a decay and its \CP conjugate. The first decay in which direct \CP violation was observed in \B mesons was {\decay{\Bd}{\Kp\pim}}~\cite{Aubert:2004qm,Chao:2004jy}. The measured \CP asymmetry of this channel is known to be ${\mathcal{A}_{\CP} = -0.082 \pm 0.006}$~\cite{PDG2018}. Decays of the \Bd meson to \pion\kaon final states and to their vector counterparts, \rhomeson\Kstar, proceed via a remarkably rich set of contributing amplitudes. For the neutral modes {\decay{\Bd}{\piz\Kz}} and \brhokst, the tree level contribution, $b \to \uubar\squark$ (depending on the CKM matrix elements $V_{ub}V^*_{us}$), is doubly Cabibbo suppressed and higher order diagrams dominate the decay (see \fig{diagrams}). Such contributions originate from the  $b \to \ddbar\squark$ ($V_{tb}V^*_{ts}$) process that may proceed either via colour-allowed electroweak-penguin or gluonic-penguin transitions. When accounting for the helicity amplitudes of the \brhokst~vector-vector ($VV$) decay, the electroweak-penguin amplitude contributes with different signs depending on the considered helicity eigenstate. This allows for several interference patterns in the decay and plays an important role in its polarisation since both penguin amplitudes are comparable in magnitude. A detailed discussion on these phenomena can be found in Ref.~\cite{Beneke:2006hg}. Other theoretical works~\cite{Gronau:2005pq} predict enhanced direct \CP-violating effects in the \brhokst~decay due to the interference with the \bomkst~decay and due to isospin-breaking consequences of this interference.
The angular analysis of $VV$ decays also gives access to T-odd triple product asymmetries (TPA), which are observables suitable for comparison with theoretical predictions, such as those in Ref.~\cite{Datta:2003mj}.

In the past, the theoretical approach to the study of \B decays into light-vector mesons was influenced by the idea that quark helicity conservation and the V$-$A nature of the weak interaction induce large longitudinal polarisation fractions, of order
$f^0\sim 0.9$. However, this prediction holds only for decays dominated by tree diagrams~\cite{Vanhoefer:2015ijw,*Aubert:2007nua}, whilst in penguin-dominated decays this hypothesis is not fulfilled~\cite{LHCb-PAPER-2017-048,LHCb-PAPER-2014-026,*Aaltonen:2011rs}.\footnote{The decay $\Bd \to \Kstarz \Kstarzb$ ($f^0= 0.80^{+0.12}_{-0.13}$~\cite{Aubert:2007xc}) seems to be an exception.
}
Low values of longitudinal polarisation fractions in penguin-dominated decays could be accounted by the SM invoking a strong-interaction effect, both in the QCD factorisation (QCDF)~\cite{Beneke:2006hg} and perturbative (pQCD)~\cite{Zou:2015iwa} frameworks.
This so-called polarisation puzzle might be resolved by combining measurements from all the {\decay{B}{\rhomeson\Kstar}} modes  (\brhokst, {\decay{\Bd}{\rhom\Kstarp}}, {\decay{\Bu}{\rhoz\Kstarp}} and {\decay{\Bu}{\rhop\Kstarz}}). This would allow also to probe physics beyond the SM~\cite{Baek:2005jk,Datta:2004jm}. 

The decay mode {\decay{\Bd}{\rhoz\Kstarz}} and its scalar-vector counterpart {\decay{\Bd}{f_0(980)\Kstarz}} have previously been studied by the \babar~\cite{Lees:2011dq} and \belle~\cite{Kyeong:2009qx} collaborations. The \babar collaboration determined the longitudinal polarisation fraction of the \CP-averaged {\decay{\Bd}{\rhoz\Kstarz}} decay to be $f^0 = 0.40 \pm 0.08 \pm 0.11$. The measurement of the \CP-averaged longitudinal polarisation of {\decay{\Bd}{{\decay{\omegaz ( }{\pipic\piz}})\Kstarz}} decays has been performed by both \babar and \belle collaborations yielding ${f^0 = 0.72 \pm 0.14 \pm 0.02}$~\cite{Aubert:2009sx} and ${f^0 = 0.56 \pm 0.29^{+0.18}_{-0.08}}$~\cite{Goldenzweig:2008sz}, respectively.

In this paper an amplitude analysis of the \Bd decay to (\pipic)(\kpic) final state in the two-body invariant mass windows ${300 < m(\pipic)<1100}\mevcc$ and ${750 < m(\kpic)<1200}\mevcc$ is presented. The analysis uses the data sample collected during the \lhc Run I, corresponding to an integrated luminosity of $1 \invfb$ of \proton\proton collisions taken by the \lhcb experiment in $2011$ at a centre-of-mass energy of $\sqs = 7 \tev$ and to $2 \invfb$ recorded during $2012$ at $\sqs = 8 \tev$. 
In the considered (\pipic) invariant-mass range the vector resonances $\rho^0$ and $\omega$ are expected to contribute, together with the scalar resonances $f_0(500)$, $f_0(980)$ and $f_0(1370)$. The (\kpic) spectrum is dominated by the vector $\Kstar(892)^0$ resonance, but contributions due to the nonresonant (\kpic) interaction and the $K^*_0(1430)^0$ state are also accounted for.
A measurement of the \CP asymmetries for the different amplitudes is made, whereas no attempt is done to measure the overall branching fraction or the global direct \CP asymmetry. The focus of the analysis is on the polarisation fractions of the vector-vector modes as well as the relative phases of the different contributions.
\section{Detector and simulation}
\label{sec:Detector}

The \lhcb detector~\cite{Alves:2008zz,LHCb-DP-2014-002} is a single-arm forward spectrometer covering the \mbox{pseudorapidity} range $2<\eta <5$, designed for the study of particles containing \bquark or \cquark quarks. The detector includes a high-precision tracking system consisting of a silicon-strip vertex detector surrounding the $pp$ interaction region, a large-area silicon-strip detector located upstream of a dipole magnet with a bending power of about $4{\mathrm{\,Tm}}$, and three stations of silicon-strip detectors and straw drift tubes placed downstream of the magnet. The tracking system provides a measurement of the momentum, \ptot, of charged particles with a relative uncertainty that varies from 0.5\% at low momentum to 1.0\% at 200\gevc. The minimum distance of a track to a primary vertex (PV), the impact parameter (IP), is measured with a resolution of $(15+29/\pt)\mum$, where \pt is the component of the momentum transverse to the beam, in\,\gevc. Different types of charged hadrons are distinguished using information from two ring-imaging Cherenkov detectors. Photons, electrons and hadrons are identified by a calorimeter system consisting of scintillating-pad and preshower detectors, an electromagnetic and a hadronic calorimeter. Muons are identified by a system composed of alternating layers of iron and multiwire proportional chambers. The identification of the particles species (PID) is performed with dedicated neural networks based on discriminating variables that combine information from the above mentioned detectors~\cite{RICH}.

The online event selection is performed by a trigger, which consists of a hardware stage, based on information from the calorimeter and muon systems, followed by a software stage, which applies a full event reconstruction. In the offline selection, trigger signals are associated with reconstructed particles. Selection requirements can therefore be made on the trigger selection itself and on whether the decision was due to the signal candidate (Triggered On Signal, \texttt{TOS}), other particles produced in the $pp$ collision (Triggered Independent of Signal, \texttt{TIS}), or a combination of both. In this work, the overlap of both trigger categories is included in the \texttt{TIS} category and candidates are split according to \texttt{TIS} and \texttt{TOSnotTIS} trigger decision to define disjoint analysis samples.

Simulated samples are used to describe the detector acceptance effects, to optimise the selection of signal candidates and to describe the \bskstkst background. They are corrected using data. Simulated samples of both resonant, \brhokst, and nonresonant, {\decay{\Bd}{(\pipic)(\kpic)}}, modes are combined to describe the signal candidates. In the simulation, $pp$ collisions are generated using \pythia~\cite{Sjostrand:2007gs,Sjostrand:2006za} with a specific \lhcb configuration~\cite{LHCb-PROC-2010-056}. Decays of hadronic particles are described by \evtgen~\cite{Lange:2001uf}, in which final-state radiation is generated using \photos~\cite{Golonka:2005pn}. The interaction of the generated particles with the detector, and its response, are implemented using the \geant toolkit~\cite{Allison:2006ve, *Agostinelli:2002hh} as described in Ref.~\cite{LHCb-PROC-2011-006}.
\section{Signal selection}
\label{sec:selection}

The event selection is based on the topology of the ${\Bd \to \rhoz(\to \pipic)\Kstarz(\to\kpic)}$ decay. Each vector-resonance candidate is formed by combining two pairs of oppositely charged tracks that are required to originate from a common vertex, to have transverse momentum above 500\mevc and large impact parameter significance, \chisqip$>16$, with respect to any PV in the event. Here the impact parameter significance is defined as the difference in the vertex fit \chisq of a given PV when it is reconstructed with and without the track candidate. In addition, each vector resonance candidate is required to have transverse momentum larger than 900\mevc and total momentum larger than 1\gevc. The \Bd candidates are formed by combining the aforementioned four tracks, which must form a good quality vertex. These candidates are required to have flight direction aligned with their momentum vector and a small significance, \chisqip$<20$, of the impact parameter with respect to their production PV.

The final-state particle with the largest neural network PID kaon hypothesis is assigned to be the kaon candidate, while the remaining three particles are required to be consistent with the pion hypothesis. A dedicated PID requirement on the kaon candidate, against its probability of being identified as a proton, reduces the contribution from the {\decay{\Lb}{\proton\pim\pip\pim}} decay mode to a negligible level. Pairs of (\pipic) and (\kpic) are formed selecting the combinations that fulfil the two-body invariant-mass range requirements ${300  < m(\pipic) < 1100}$\mevcc and ${750 < m(\kpic) < 1200}$\mevcc, while having a four-body invariant mass within the ${5190 < m(\pipic \kpic) < 5700}$\mevcc range. Backgrounds from partially reconstructed \Bd decays do not enter the selected $m(\pipic\kpic)$ invariant-mass range. The potential ambiguity on the assignment of the same-sign pions to the (\pipic) and (\kpic) pairs is reduced to a negligible level by the requirements on the invariant masses. 

A possible source of background is due to {\decay{\Bd}{{\decay{\Dzb(}{\kpic})}\pipic}} decays, where the final state particles are incorrectly paired. To remove this background, candidates are reconstructed under the alternate pairing hypothesis and those within a 20\mevcc window around the known mass of the \Dz meson~\cite{PDG2018} are rejected. The requirements placed on the two-body invariant masses and a dedicated constraint on one of the angular variables, $|\cos\theta_{\pipin}| <0.8$ (variable defined in \refsec{ampfit}), strongly suppress background contributions from other decays proceeding via three-body resonances, such as {\decay{\Bd}{\Dm\pip}}, {\decay{\Bd}{a_1(1260)^-\Kp}} or {\decay{\Bd}{K_1(1270)^+\pim}}.

Background due to random combinations of tracks (combinatorial) is suppressed by means of a boosted decision tree (BDT)~\cite{Breiman,AdaBoost} multivariate classifier. The discriminating power of the BDT is achieved using several kinematic (transverse momentum of the \Bd candidate) and topological variables (related to the \Bd decay vertex, such as the fit quality and the separation from the PV), which are optimal for discrimination between the signal and the background while not biasing the two-body invariant mass distributions. 

Different BDTs are trained for the 2011 and 2012 data-taking periods to account for their different centre-of-mass energies. Candidates in the upper side band of the four-body invariant mass spectrum, $m(\pipic \kpic)>5540$\mevcc, are used as the background training sample while candidates from a simulated signal sample are used as the signal training sample. Both samples are randomly split into two to allow for testing of the BDT performance and to check for possible overtraining of the algorithm. The optimal threshold for the BDT output value is determined by requiring a background rejection power in the training and testing samples larger than 99\%. This choice maximises the product of signal purity and significance. Once the full selection is applied, 0.1\% of events contain multiple candidates, which share at least one of the final-state particles. Among these, only the candidate with the highest BDT output value is kept in the analysed sample. The resulting data sample is dominated by signal candidates, with a small contribution from random combinations of tracks and from \bskpikpi decays.
In addition, there is a hint of a {\decay{\Bsb}{(\pipic)(\kpic)}} contribution in the selected data sample.

\section{Fit to the four-body invariant mass spectrum}
\label{sec:four-body-spec}
A fit to the four-body invariant mass distribution is performed simultaneously on the four categories studied in the analysis (split according to trigger decision and data-taking year). The fit is also simultaneous in the two charge-conjugate final states, which define the \Bd and \Bdb samples. 

For each category, signal weights from the four-body invariant mass fit are used to produce background-subtracted data samples by means of the \sPlot~\cite{Pivk:2004ty} technique. This allows the amplitude fit to be performed on a sample that represents only the signal and avoids making assumptions on the multidimensional shapes of the backgrounds.

Prior to performing the four-body invariant-mass fit, the \bskpikpi~contribution is subtracted by injecting simulated events with negative weights after estimation of their per-category yield. In order to perform this estimation, the PID selection requirement on one of the final-state pions is changed to select (\kpic)(\kpim) candidates instead of the nominal (\pipic)(\kpic) final state. The (\kpic)(\kpim) four-body invariant-mass spectrum is fitted to obtain the yield of this background, which is then corrected by the ratio of PID efficiencies, computed using data, to obtain its final contribution to the analysed data sample. The reason for this particular treatment is that, when the kaon is misidentified as a pion, the reconstructed mass of these candidates spans widely in the spectrum underneath the \Bd and \Bsb signal peaks. To ensure a proper cancellation of this background, the injected \bskpikpi~simulated events are weighted according to a probability density function (PDF) whose physical parameters (describing the $VV$, $VS$ and $SS$ amplitudes) are taken from a previous measurement~\cite{LHCb-PAPER-2014-068}.

\begin{figure}[t]
\begin{centering}
\includegraphics[width=0.495\textwidth]{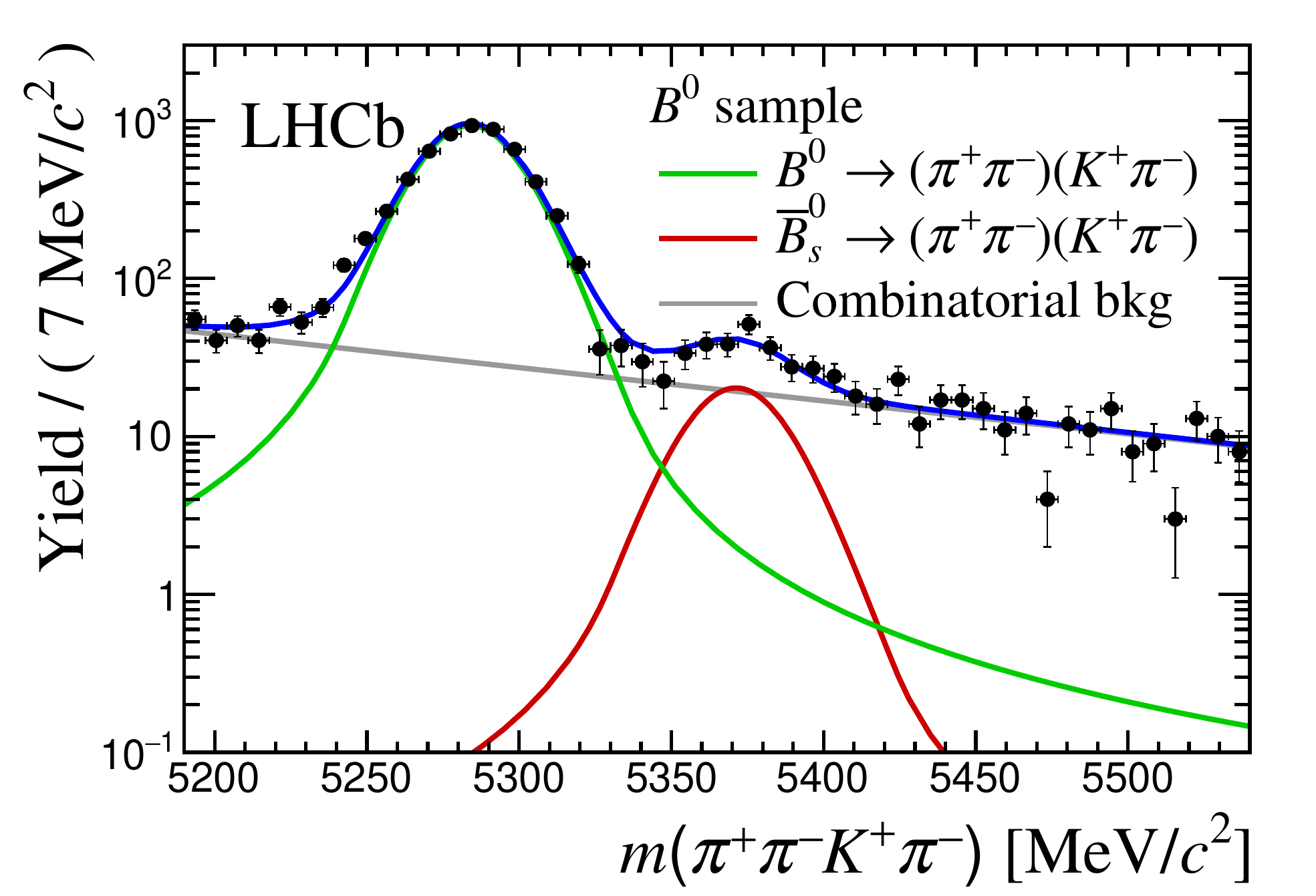}
\includegraphics[width=0.495\textwidth]{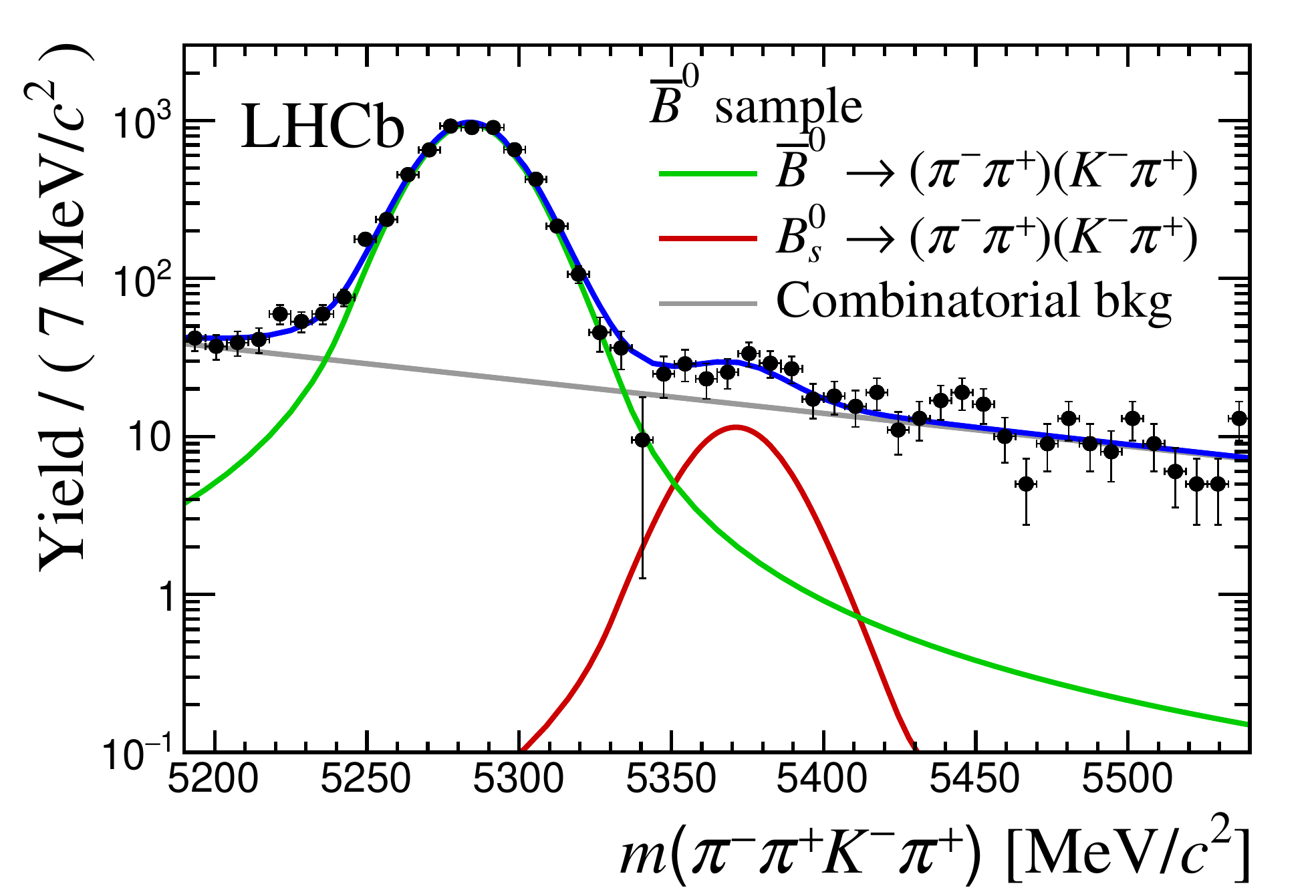}
\caption{\small Fit to the invariant-mass distribution of selected (left) \Bd and (right) \Bdb candidates after the subtraction of \bskstkst background decays. The four trigger and data-taking year categories are aggregated in the figures. The contributions due to the \decay{\Bd}{(\pipic)(\kpic)} signal, \decay{\Bsb}{(\pipic)(\kpic)} background and combinatorial background are represented by the solid green, red and grey lines, respectively. Data are shown using black dots and the overall fit is represented by the solid blue line.}
\label{fig:4bodyfit}
\end{centering}
\end{figure}

The resulting data samples are fitted to a model where the signal peak is described with an Hypatia distribution~\cite{Santos:2013gra}, consisting of a Gaussian-like core and asymmetric tails. Its parameters, except for the mean and width values which are free to vary, are determined from a fit to the distribution of signal candidates obtained from simulation. The contribution from the {\decay{\Bsb}{(\pipic)(\kpic)}} mode is described by the same distribution used for the signal, except for its mean value that is shifted by the known \Bs and \Bd mass difference~\cite{PDG2018}. Finally, an exponential function accounts for the combinatorial background. Figure~\ref{fig:4bodyfit} shows the simultaneous four-body invariant-mass fit result separated for \Bd and \Bdb samples. \tab{4bodyfit} shows the yields obtained in each of the eight fitting categories.

\begin{table}[t!]
\caption{\small{Yields obtained in the extended simultaneous four-body invariant mass fit to the four categories and for the two final states. The quoted uncertainties are statistical only.}}
\label{tab:4bodyfit}
\begin{center}
\begin{tabular}{c|c|c|ccc}
\toprule
Final State & Year & Trigger & \Bd & \Bsb & Combinatorial \\
\midrule
\multirow{4}{*}{(\pipic)(\kpic)} & \multirow{2}{*}{2011} & \texttt{TIS} & $\phantom{0}985 \pm 34$ & $\phantom{0}20 \pm \phantom{0}9$ & $\phantom{0}249 \pm 23$ \\
 & & \texttt{TOSnoTIS} & $\phantom{0}615 \pm 27$ & $\phantom{00}7 \pm \phantom{0}5$ & $\phantom{0}134 \pm 17$ \\
\\[-0.85em]
 \cdashline{2-6}
 \\[-0.85em]
 & \multirow{2}{*}{2012} & \texttt{TIS} & $2451 \pm 54$ & $\phantom{0}62 \pm 13$ & $\phantom{0}487 \pm 35$ \\
 & & \texttt{TOSnoTIS} & $1422 \pm 41$ & $\phantom{0}30 \pm \phantom{0}9$ & $\phantom{0}250 \pm 24$ \\
\toprule
Final State & Year & Trigger & \Bdb & \Bs & Combinatorial \\
\midrule
 \multirow{4}{*}{(\pipic)(\kpim)} & \multirow{2}{*}{2011} & \texttt{TIS} & $1013 \pm 34$ & $\phantom{00}4 \pm \phantom{0}7$ & $\phantom{0}204 \pm 22$ \\
 & & \texttt{TOSnoTIS} & $\phantom{0}620 \pm 26$ & $\phantom{00}6 \pm \phantom{0}4$ & $\phantom{00}69 \pm 12$ \\
 \\[-0.85em]
 \cdashline{2-6}
 \\[-0.85em]
 & \multirow{2}{*}{2012} & \texttt{TIS} & $2521 \pm 53$ & $\phantom{0}46 \pm 13$ & $\phantom{0}437 \pm 32$ \\
 & & \texttt{TOSnoTIS} & $1439 \pm 40$ & $\phantom{0}12 \pm \phantom{0}7$ & $\phantom{0}220 \pm 23$ \\
 \bottomrule
\end{tabular}

\end{center}
\end{table}
\section{Amplitude fit}
\label{sec:ampfit}
An amplitude analysis is performed on the background-subtracted samples obtained as described in~\refsec{four-body-spec}. The isobar model\cite{Fleming:1964zz,Morgan:1968zza,Herndon:1973yn}, in which an overall rate is built from the coherent sum over the considered contributions, is used to build the total decay amplitude under the quasi-two-body assumption. In the nominal fit, a total of fourteen components, listed in~\tab{amps}, are accounted for in the analysed region of the (\pipic) and (\kpic) two-body invariant masses. 

\begin{figure}
\begin{centering}
\includegraphics[width=0.7\textwidth]{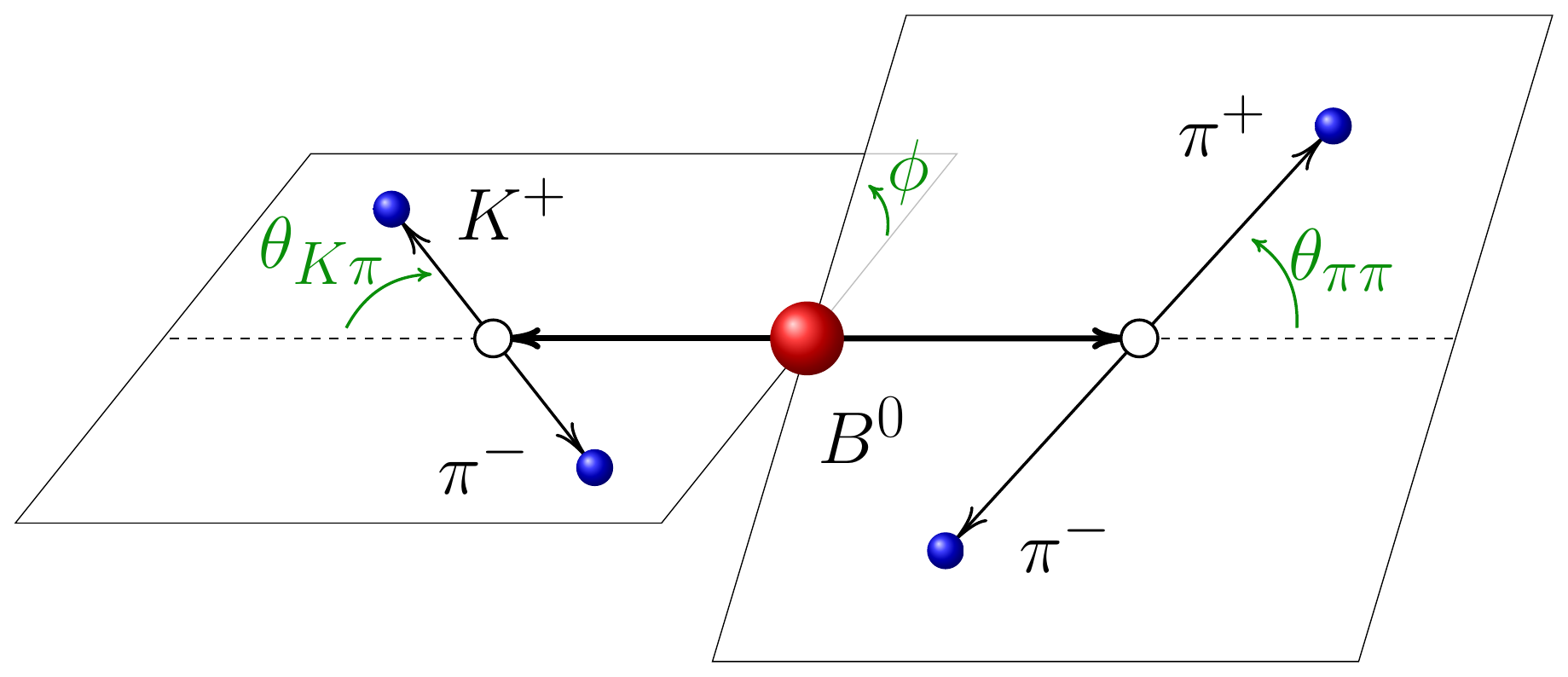}
\caption{\small{Definition of the helicity angles in the \brhokst decay.}}
\label{fig:VVangles}
\end{centering}
\end{figure}

The angular distributions are described using the helicity angles, depicted in~\fig{VVangles}, where $\theta_{\pipin}$ is the angle between the \pip direction in the (\pipic) rest frame and the (\pipic) direction in the \Bd rest frame,
$\theta_{\kpin}$ is the angle between the \Kp direction in the (\kpic) rest frame and the (\kpic) direction in the \Bd rest frame, and $\phi$ is the angle between the (\pipic) and the (\kpic) decay planes. The angular functions, ${g_i(\theta_{\pipin},\theta_{\kpin},\phi)}$, are built from spherical harmonics and are listed in~\tab{amps}. The dependence of the total amplitude on the two-body invariant masses, $R_i(m_{\pipin},m_{\kpin})$, is described by the product of (\pipic) and (\kpic) propagators, $M(m_{ij})$, and distinguishes resonances with the same angular dependence. These terms depend on the mass propagator choice and are described as
\begin{equation}
\label{eq:reslineshape}
\begin{split}
R_i(m_{\pipin},m_{\kpin}) = B_{L_{B^0}}\times \left(\frac{q_{(K\pi)(\pi\pi)}}{m_{B^0}}\right)^{L_{B^0}} 
 \times B_{L_R}\times\left(\frac{q_{\pipin}}{m_R}\right)^{L_R} \times M(m_{\pipin}) \\
 \times B_{L_{R'}}\times\left(\frac{q_{\kpin}}{m_{R'}}\right)^{L_{R'}} \times M'(m_{\kpin})
 \times \Phi(m_{\pipin},m_{\kpin}),
\end{split}
\end{equation}
where $q_{ij}$ stands for the relative momentum of the final state particles in their parent's rest frame; $\Phi(m_{\pipin},m_{\kpin})$ represents the four-body phase-space density, $m_{R^{(')}}$, the Breit--Wigner mass of the resonance $R^{(')}$; and $B_L$ represents the Blatt--Weisskopf \cite{Blatt:1952ije} barrier penetration factor, which depends on the resonance radius and on the relative angular momentum between the decay products, $L$. The value of $L$ influences not only the angular distributions but also the shapes of the two-body invariant-mass distributions due to the aforementioned barrier factors, which originate in the production and decay processes of a resonance. In the nominal fit the barrier factor arising from the production process of the vector-mesons is not included, and thus the value $L_{B^0}=0$ is used. A systematic uncertainty is assigned because of this assumption.

In the selected region of (\pipic) invariant mass the following resonances are expected to contribute and thus are included. The scalar ($S$) resonances \fzfh and \fzhm, described with relativistic spin-0 Breit--Wigner functions, and the \fznh meson, described with a Flatt\'e parametrisation~\cite{flatte,*Flatte2}. 
Also included are the vector ($V$) resonances \omegaz, described with a relativistic spin-1 Breit--Wigner shape, and \rhoz, described with the Gounaris--Sakurai parametrisation~\cite{Gounaris-Sakurai}. The functional forms of these parametrisations are given in~\app{massprop}.

The analysed invariant mass region of (\kpic) candidates is dominated by two contributions: the vector $K^*(892)^0$ resonance, described with a relativistic spin-1 Breit--Wigner, and scalar states, which are comprised of the resonant state $K^*_0(1430)^0$ and a nonresonant component. The phase evolution of the scalar amplitude is parametrised by the LASS function~\cite{lass}, while its modulus is modified with a real exponential form factor obtained from a one-dimensional fit to the (\kpic) invariant-mass spectrum of the efficiency-corrected data sample. 

Depending on the spin of the resonant states, different possible amplitudes can contribute to the final state: the combination of two scalars or of a scalar with a vector resonance proceeds via one possible configuration, while in case of two vector resonances three transversity amplitudes contribute to the decay rate ($A^0, A^{||}$ and $A^{\perp}$). The transversity ($0,||,\perp$) basis is obtained from a linear transformation of the helicity ($00, ++,--$) states that results in amplitudes with defined \P eigenvalues. \tab{amps} gathers the list of considered amplitudes with their corresponding parity and the angular and two-body invariant mass dependence for each term.

The fit PDF is defined as the differential decay rate,
\begin{align}
\label{eq:DecRate}
&\frac{{\rm d}^5\Gamma}{{\rm d}m^2_{\pipin}{\rm d}m^2_{\kpin}{\rm d}\cos\theta_{\pipin}{\rm d}\cos\theta_{\kpin}{\rm d}\phi} \propto
\Phi(m_{\pipin},m_{\kpin}) \times \\ &\phantom{00000} \sum^{14}_{i=1}\sum^{14}_{j=1}[A_i  g_i(\theta_{\pipin},\theta_{\kpin},\phi)  R_i(m_{\pipin},m_{\kpin})] [A_j  g_j(\theta_{\pipin},\theta_{\kpin},\phi)  R_j(m_{\pipin},m_{\kpin})]^*, \nonumber
\end{align}
where indices $i$ and $j$ run over the list given by the first column of~\tab{amps}. 

\begin{table}[t]
\caption{Contributions to the total amplitude and their angular and mass dependencies.}
\label{tab:amps}
\begin{center}
\resizebox{\textwidth}{!}{
\renewcommand{\arraystretch}{1.4}
\begin{tabular}{c c c l c  c}
\toprule
$i$ & State & Parity & $A_i$ & $g_i (\theta_{\pipin},\theta_{\kpin},\phi)$ & $M(m_{\pipin})M(m_{\kpin})$ \\
\midrule
1 & $VV$ &$\phantom{-}1$ & $A^0_{\rhomeson\Kstar}$ & $\cos\theta_{\pipin}\cos\theta_{\kpin} $&$ M_{\rhomeson}(m_{\pipin})M_{\Kstar}(m_{\kpin}) $  \\
2 & $VV$ & $\phantom{-}1$ & $A^{||}_{\rhomeson\Kstar}$ & $\frac{1}{\sqrt 2} \sin\theta_{\pipin}\sin\theta_{\kpin}\cos\phi $&$ M_{\rhomeson}(m_{\pipin})M_{\Kstar}(m_{\kpin})  $\\
3 & $VV$ & $-1$ & $A^{\perp}_{\rhomeson\Kstar}$ & $\frac{i}{\sqrt 2} \sin\theta_{\pipin}\sin\theta_{\kpin}\sin\phi $&$ M_{\rhomeson}(m_{\pipin})M_{\Kstar}(m_{\kpin}) $ \\
\midrule
4 & $VV$ & $\phantom{-}1$ & $A^0_{\omegaz\Kstar}$ & $\cos\theta_{\pipin}\cos\theta_{\kpin} $&$ M_\omegaz(m_{\pipin})M_{\Kstar}(m_{\kpin})   $\\
5 & $VV$ & $\phantom{-}1$ & $A^{||}_{\omegaz\Kstar}$ & $\frac{1}{\sqrt 2} \sin\theta_{\pipin}\sin\theta_{\kpin}\cos\phi $&$ M_\omegaz(m_{\pipin})M_{\Kstar}(m_{\kpin})  $\\
6 & $VV$ & $-1$ & $A^{\perp}_{\omegaz\Kstar}$ & $\frac{i}{\sqrt 2} \sin\theta_{\pipin}\sin\theta_{\kpin}\sin\phi $&$ M_\omegaz(m_{\pipin})M_{\Kstar}(m_{\kpin}) $ \\
\midrule
7 & $VS$ &$\phantom{-}1$ & $A_{\rhomeson (\kaon\pion)}$ & $\frac{1}{\sqrt 3}\cos\theta_{\pipin} $&$ M_{\rhomeson}(m_{\pipin})M_{(K\pi)}(m_{\kpin})   $\\
8 & $VS$ &$\phantom{-}1$ & $A_{\omegaz(\kaon\pion)}$ & $\frac{1}{\sqrt 3}\cos\theta_{\pipin} $&$ M_\omegaz(m_{\pipin})M_{(K\pi)}(m_{\kpin})  $ \\
\midrule
9 & $SV$ &$\phantom{-}1$ & $A_{f_0(500)\Kstar}$ & $\frac{1}{\sqrt 3}\cos\theta_{\kpin} $&$ M_{f_0(500)}(m_{\pipin})M_{\Kstar}(m_{\kpin})   $\\
10 & $SV$ &$\phantom{-}1$ & $A_{f_0(980)\Kstar}$ & $\frac{1}{\sqrt 3}\cos\theta_{\kpin} $&$ M_{f_0(980)}(m_{\pipin})M_{\Kstar}(m_{\kpin})   $\\
11 & $SV$ &$\phantom{-}1$ & $A_{f_0(1370)\Kstar}$ & $\frac{1}{\sqrt 3}\cos\theta_{\kpin} $&$ M_{f_0(1370)}(m_{\pipin})M_{\Kstar}(m_{\kpin})   $\\
\midrule
12 & $SS$ &$\phantom{-}1$ & $A_{f_0(500) (\kaon\pion)}$ & $\frac{1}{3} $&$ M_{f_0(500)}(m_{\pipin})M_{(K\pi)}(m_{\kpin})   $\\
13 & $SS$ &$\phantom{-}1$ & $A_{f_0(980) (\kaon\pion)}$ & $\frac{1}{3} $&$ M_{f_0(980)}(m_{\pipin})M_{(K\pi)}(m_{\kpin})   $\\
14 & $SS$ & $\phantom{-}1$ & $A_{f_0(1370) (\kaon\pion)}$ & $\frac{1}{3} $&$ M_{f_0(1370)}(m_{\pipin})M_{(K\pi)}(m_{\kpin})   $\\
\bottomrule
\end{tabular}
}
\end{center}
\end{table}
The normalisation of the PDF implies that one of these quantities must be fixed to a reference value. 
For convenience, each amplitude is described in the fit by two parameters representing the real and imaginary parts. The cartesian representation of these complex quantities is preferred to avoid degeneracies in the determination of the phases in case of amplitudes with small magnitudes.
The $A_{\rhomeson(\kpin)} (VS$) component has a sizeable fit fraction, so it is picked as the reference for the normalisation of the PDF in both \Bd and \Bdb models, which  is ensured by the following arbitrary choice
\begin{equation}
\label{eq:modnorm}
{\Real}(A_{\rhomeson(\kpin)}) = 2, \;\; {\rm{and}} \;\; {\Imag}(A_{\rhomeson(\kpin)}) = 0.
\end{equation}

Therefore, the parameters that are determined from the fit correspond to the relative strength of each contribution to the decay rate with respect to that of the $VS (\rhomeson(\kpin))$, adding two degrees of freedom per contribution. To allow the identification of the squared amplitudes with the contribution of each component, relative to the $A_{\rhomeson(\kpin)}$, in the selected mass range, the mass terms are normalised according to
\begin{equation}
\label{eq:massnorm}
\int_{m'_l}^{m'_u} \int_{m_l}^{m_u} |R_i(m_{\pipin},m_{\kpin})|^2\Phi(m_{\pipin},m_{\kpin}){\rm d}m_{\pipin}^2{\rm d}m_{\kpin}^2 = 1\, ,
\end{equation}
where $m_l$ and $m_u$ are the lower and upper limits of the two-body invariant mass spectra defined in \refsec{selection}. The global phases in the considered mass propagators are arbitrarily shifted to be zero at the Breit--Wigner masses of the \rhoz and \Kstarz mesons for $m_{\pipin}$ and $m_{\kpin}$, respectively. In this way all phases are measured with respect to the same reference.

The analysed distributions are affected by the selection requirements and the detector acceptance. These effects are accounted for using the normalisation weights~\cite{duPree:1299931}, $w_{ij}$,
\begin{align}
w_{ij} = \int & \epsilon(m_{\pipin},m_{\kpin},\theta_{\pipin},\theta_{\kpin},\phi) \Phi(m_{\pipin},m_{\kpin})[g_i(\theta_{\pipin},\theta_{\kpin},\phi) R_i(m_{\pipin},m_{\kpin})]\,\times \nonumber
\\
& [g_j(\theta_{\pipin},\theta_{\kpin},\phi) R_j(m_{\pipin},m_{\kpin})]^* {\rm d}m^2_{\pipin}{\rm d}m^2_{\kpin}{\rm d}\cos\theta_{\pipin}{\rm d}\cos\theta_{\kpin}{\rm d}\phi\, , 
\label{eq:nws}
\end{align} 
\noindent 
where $\epsilon$ is the total efficiency evaluated using simulation and the $i$ and $j$ indices correspond to those of \eq{DecRate}. Since the efficiency depends on the trigger category and on the kinematics of the final-state particles, a different set of normalisation weights is calculated for each category.

From the amplitudes $A_i$, modelling \Bd decays, and $\overline{A}_i$, describing \Bdb decays, other physically meaningful observables can be derived. In particular, for the $VV$ decays \brhokst and \bomkst, these quantities are the polarisation fractions 
\begin{equation}
f_{VV}^{\lambda}=\frac{|A_{VV}^{\lambda}|^2}{|A_{VV}^0|^2+|A_{VV}^{||}|^2+|A_{VV}^{\perp}|^2}\, , \qquad \lambda = 0,||,\perp\,
\end{equation}
with their \CP averages, $\tilde{f}$, and asymmetries, $\mathcal{A}$,
\begin{equation}
\tilde{f}_{VV}^{\lambda}=
\frac{1}{2}(f_{VV}^{\lambda}+\overline{f}_{VV}^{\lambda})\, , \qquad
\mathcal{A}_{VV}^{\lambda}=
\frac{\overline{f}_{VV}^{\lambda}-f_{VV}^{\lambda}}{\overline{f}_{VV}^{\lambda}+f_{VV}^{\lambda}}\, ,
\end{equation}
and the phase differences, measured with respect to the reference channel, \mbox{{\decay{\Bd}{\rhoz(\kpin)}}},
\begin{equation}
    \delta^0_{VV} \equiv (\delta^0_{VV} - \delta_{\rho(\kpin)}) = \arg(A_{VV}^0/A_{\rho(\kpin)}).
\end{equation}
For comparison with theoretical predictions it is also convenient to compute the phase differences among the different $VV$ amplitudes, 
\begin{equation}
\delta_{VV}^{||-0,\perp-0}\equiv (\delta^{||,\perp}_{VV} - \delta^0_{VV})=\arg(A_{VV}^{||,\perp}/A_{VV}^0).
\end{equation}
From these sets of observables, the phase differences of the \CP average, $\frac{1}{2}(\delta_{\Bbar}+\delta_B$), and \CP difference, $\frac{1}{2}(\delta_{\Bbar}-\delta_B$), are obtained. Ambiguities in this definition are resolved by choosing the smallest value of the \CP-violating phase.

Finally, T-odd quantities as defined in Ref.~\cite{Datta:2003mj} can be obtained from combinations of the polarisation fractions and their phase differences as
\begin{equation}
\label{eq:TP2}
\mathcal{A}_{\text{T}}^1 = f_\perp f_0 \sin(\delta_\perp - \delta_0)\, ,
\quad
\mathcal{A}_{\text{T}}^2 = f_\perp f_{||} \sin(\delta_\perp - \delta_{||})\, .
\end{equation}
The so-called \textit{true} and \textit{fake} TPA are then calculated as
\begin{equation}
\label{eq:TPA}
\mathcal{A}^k_{\text{T-\it{true}}} = \frac{\mathcal{A}_{\text{T}}^k -\mathcal{\overline{A}}^k_{\text{T}}}{2}\, ,
\quad
\quad
\mathcal{A}^k_{\text{T-\it{fake}}} = \frac{\mathcal{A}_{\text{T}}^k +\mathcal{\overline{A}}^k_{\text{T}}}{2}\, ,
\end{equation}
where $k = 1,2$ and the \textit{true} or \textit{fake} labels refer to whether the asymmetry is due to a real \CP asymmetry or due to effects from final-state interactions that are \CP symmetric. Observing a TPA value consistent with zero would not rule out the presence of \CP-violating effects, since negligible \CP averaged phase differences would suppress the asymmetries.
\section{Results}
\label{sec:results}

The nominal fit, simultaneous in eight categories, is computationally very expensive due to the high dimensionality of the model and the large number of free parameters. To cope with this issue, the PDF is computed in parallel on a Graphical Processing Unit (GPU) using the \texttt{Ipanema}~\cite{Ipanema} framework. This parallelisation reduces the computing time by a factor $\sim50$
when using \texttt{Minuit} \cite{James:1975dr} to minimise the likelihood function.   The \texttt{Ipanema} framework is implemented using pyCUDA~\cite{KLOCKNER2012157} and serves as interface to minimisation algorithms other than \texttt{Minuit}. In particular, it allows to use the \texttt{MultiNest} algorithm~\cite{multinest1,*multinest2,*multinest3}, which employs a multimodal nested sampling strategy to calculate the most likely values of the fitted parameters. Not relying on partial derivatives of the minimised function, the \texttt{MultiNest} method is very effective in finding minima of the likelihood function in weighted data samples, like in this work, and is thus preferred to \texttt{Minuit} to obtain the central values of the result. Despite its robustness, \texttt{MultiNest} is much slower than \texttt{Minuit} and therefore the latter was used to evaluate some systematic uncertainties using pseudoexperiments, as explained in~\refsec{systs}.

\begin{figure}[!ht]
\begin{centering}\def\ww{0.355\textwidth}
\includegraphics[width=\ww]{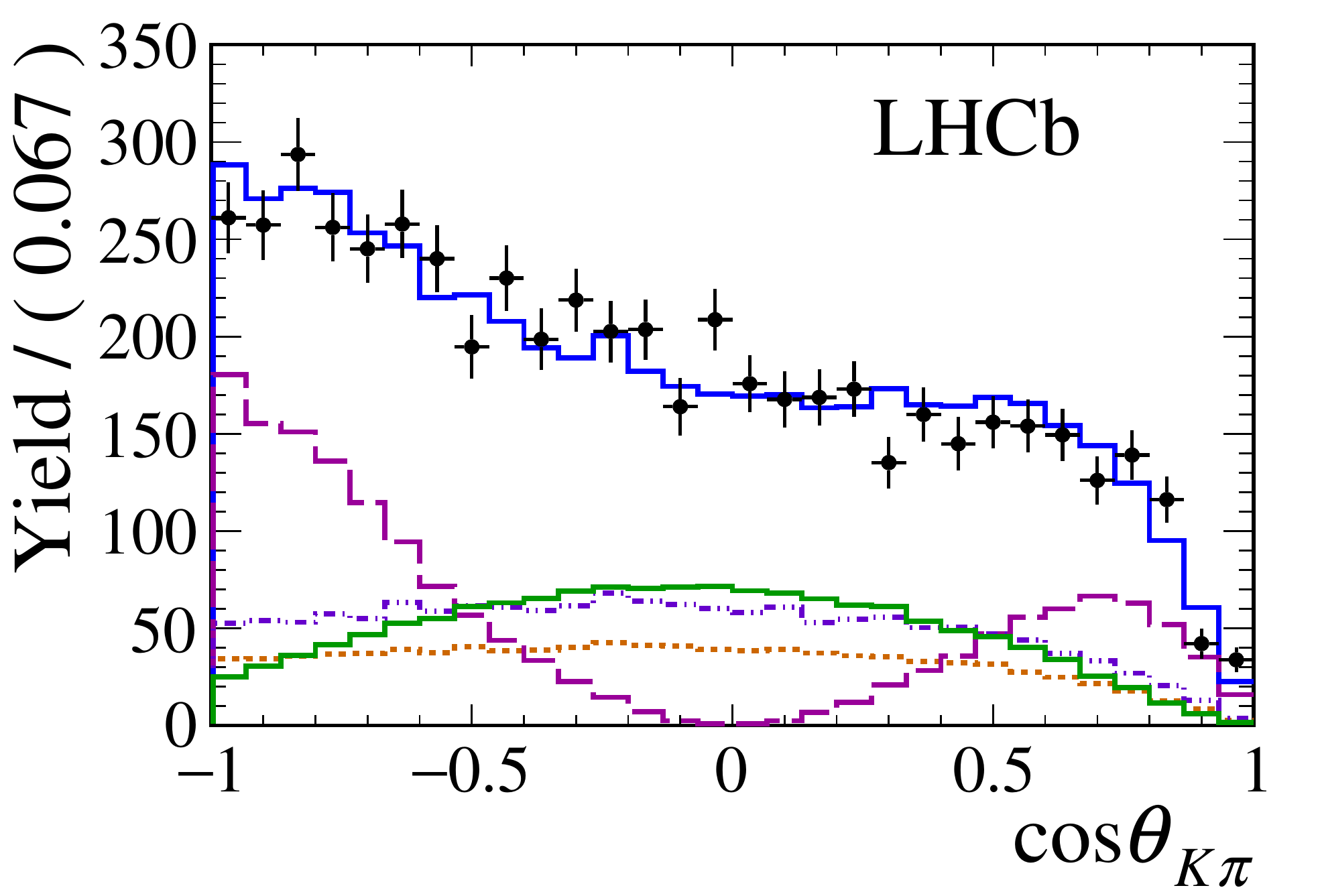}
\includegraphics[width=\ww]{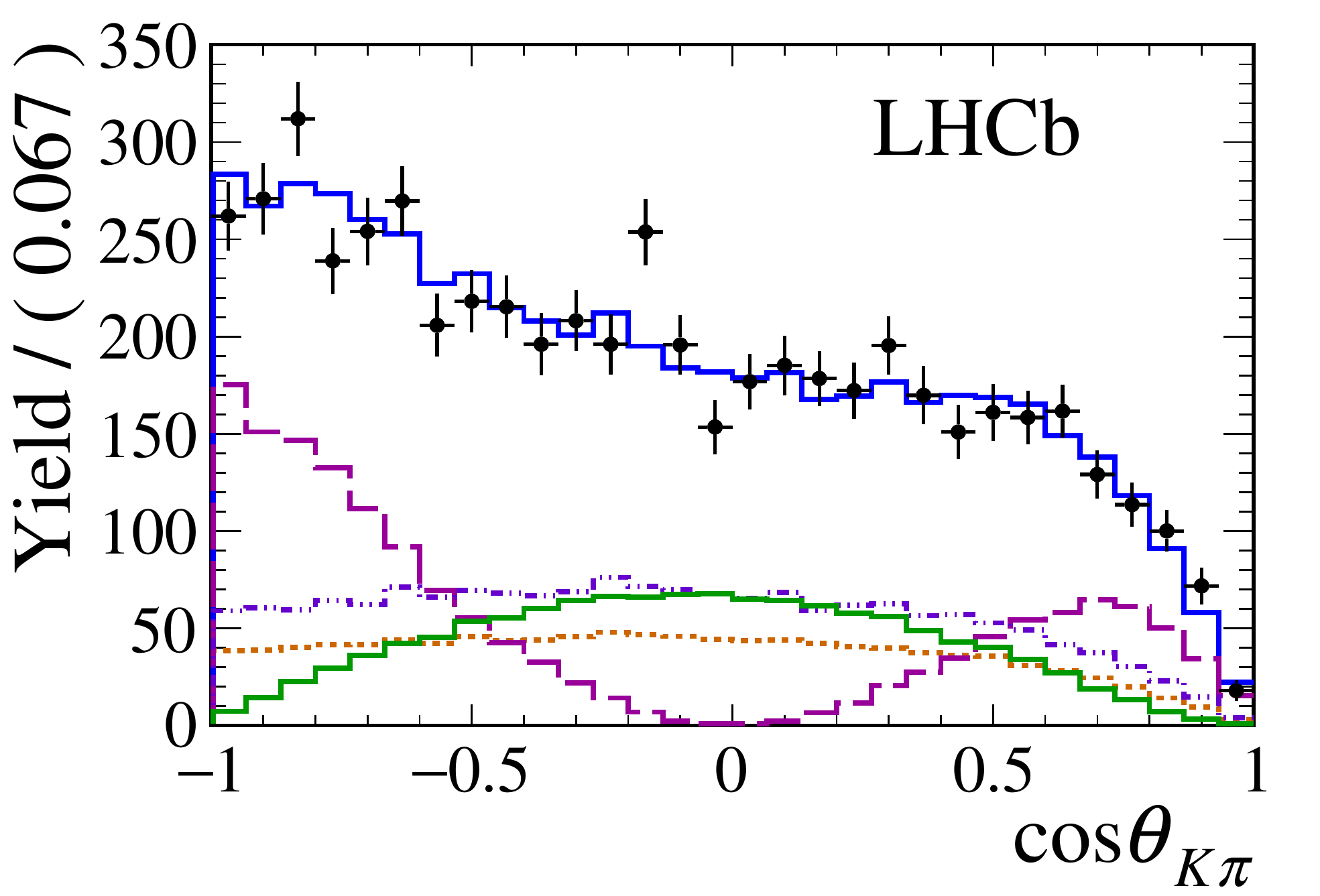}\\
\includegraphics[width=\ww]{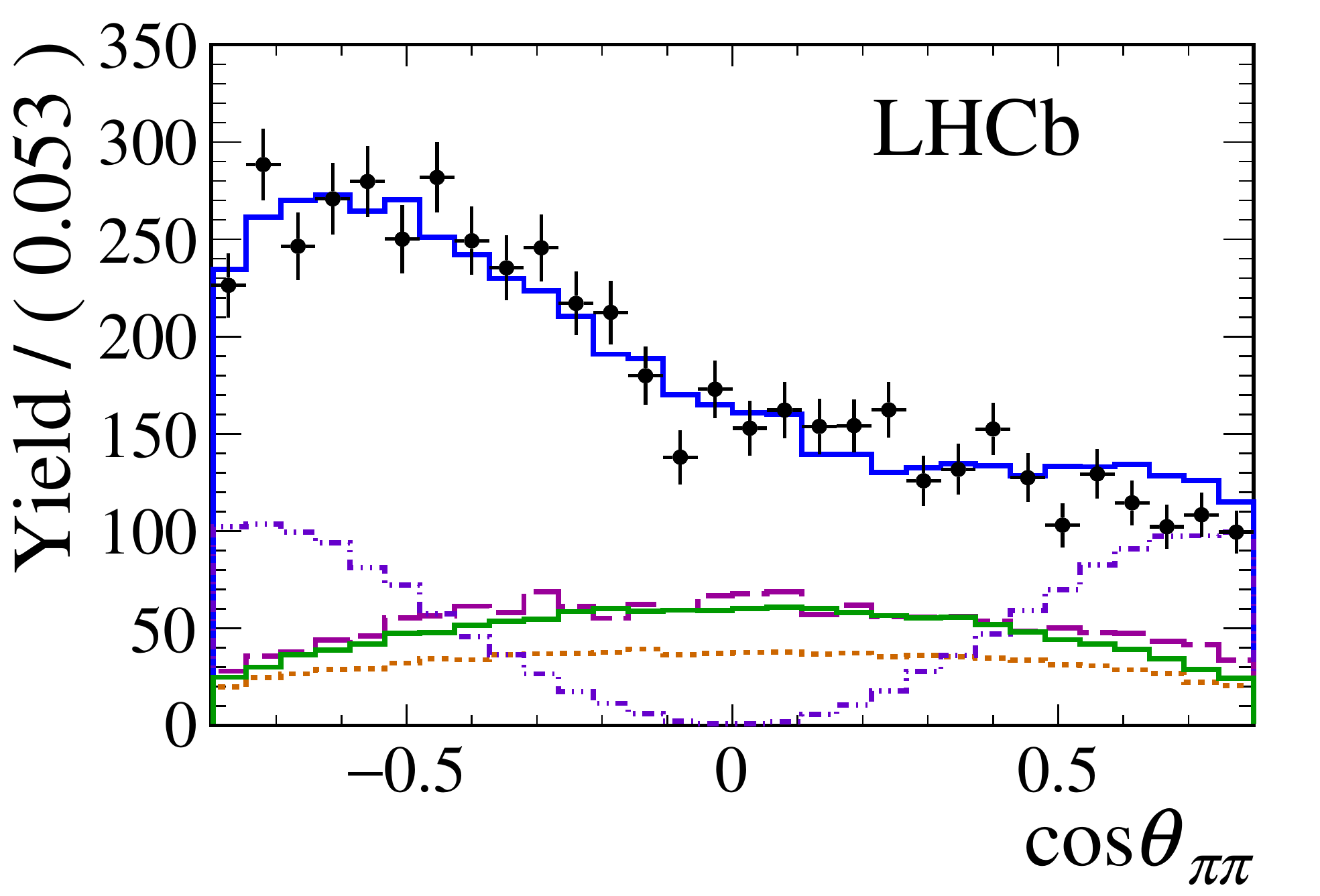}
\includegraphics[width=\ww]{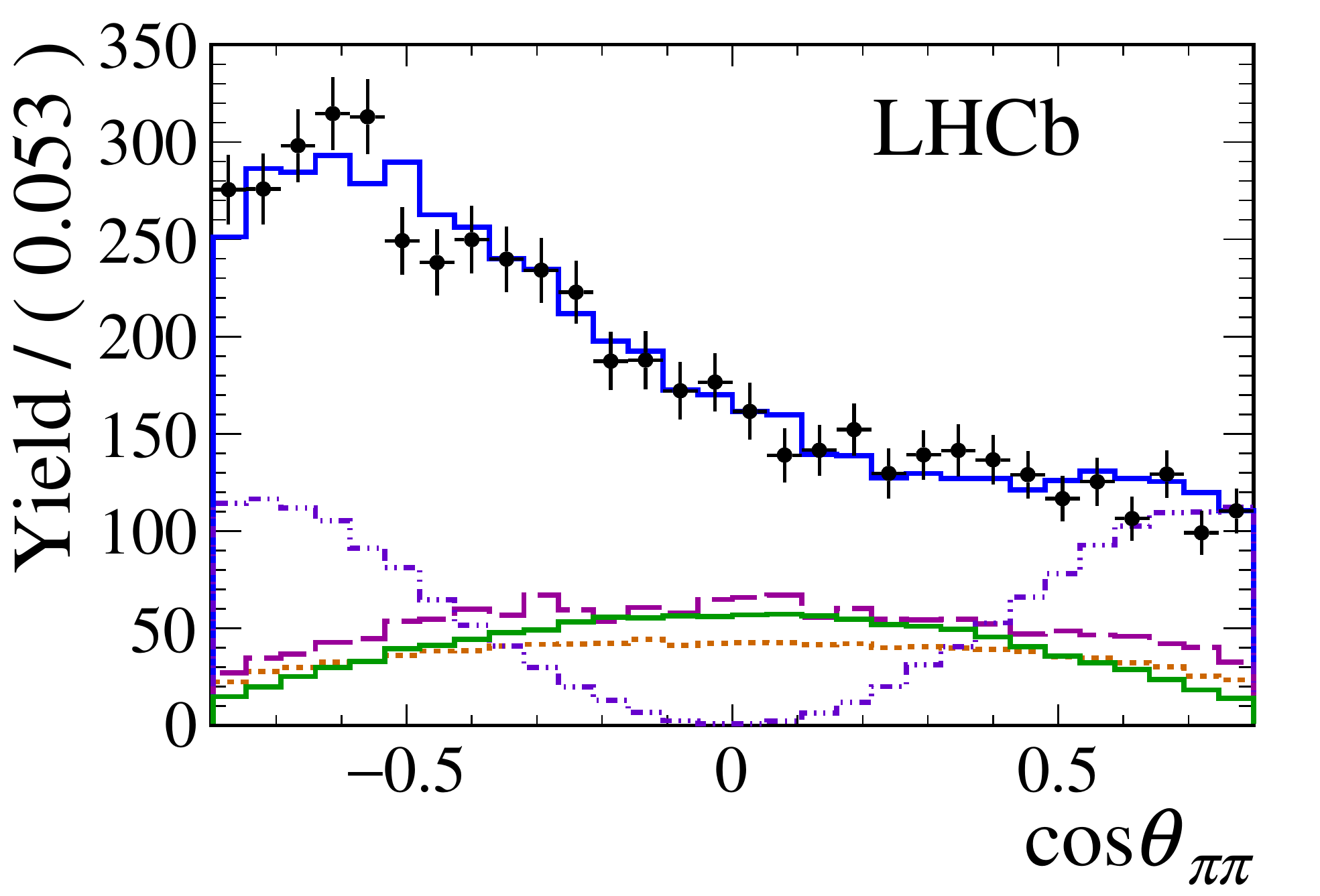}\\
\includegraphics[width=\ww]{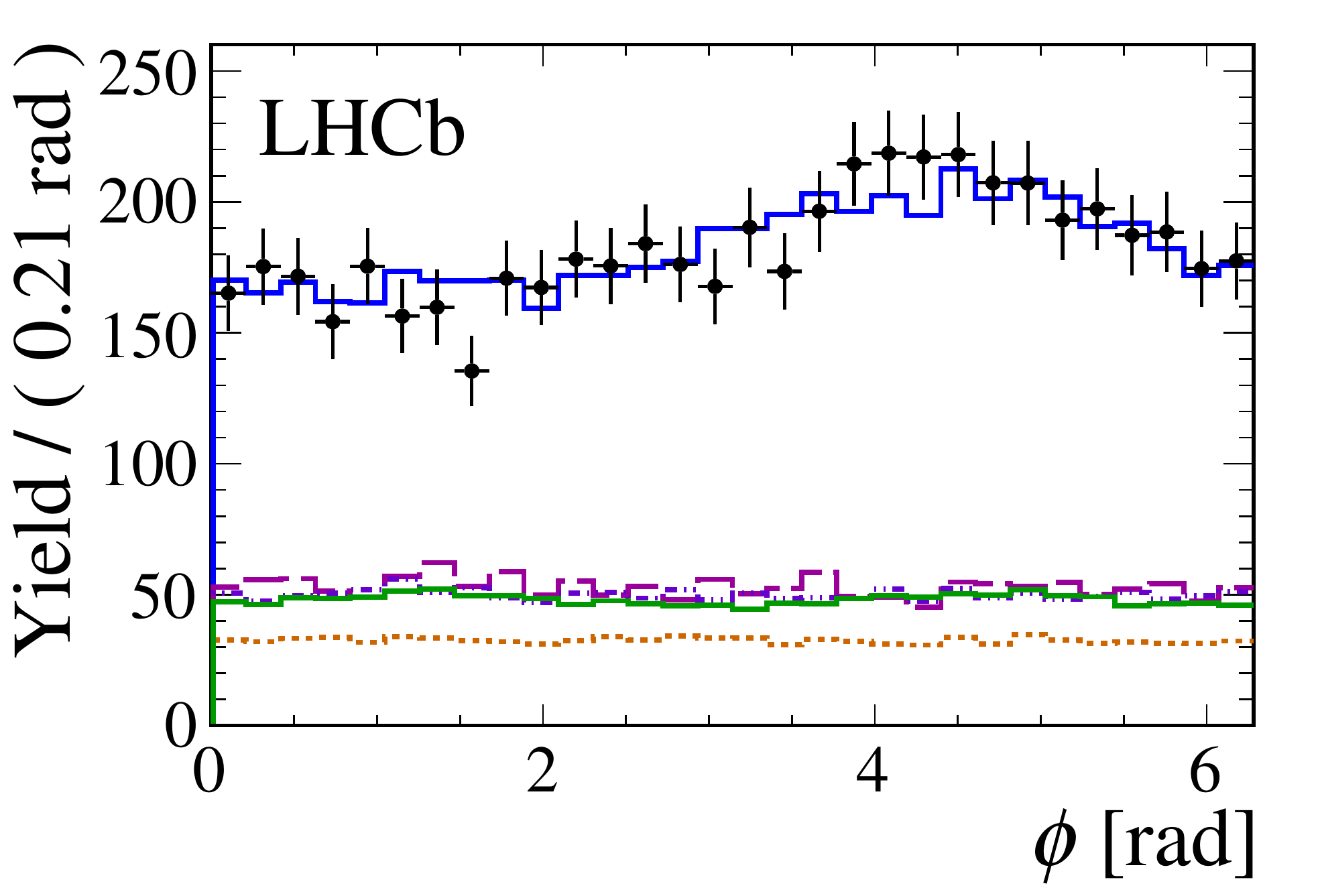}
\includegraphics[width=\ww]{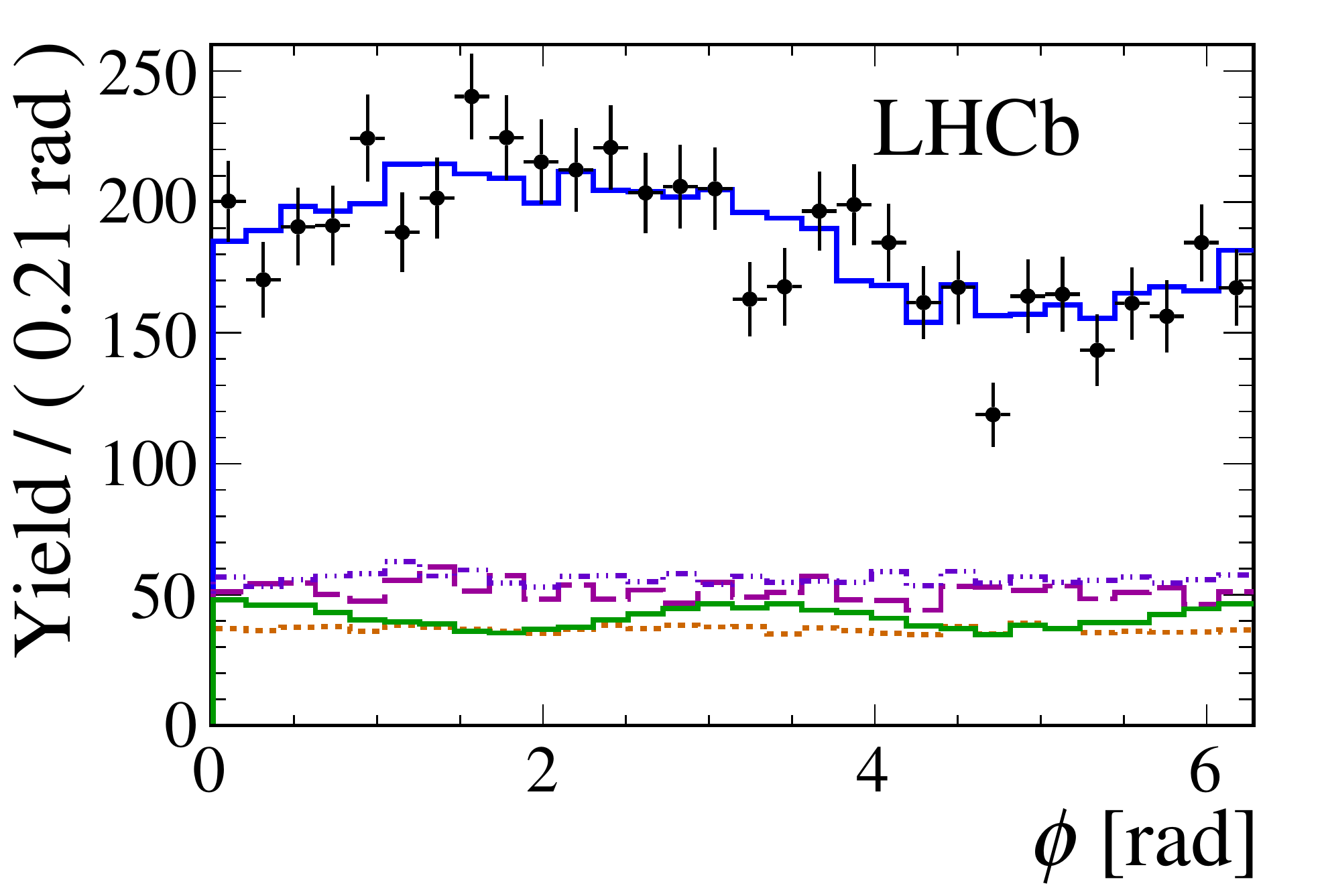}\\
\includegraphics[width=\ww]{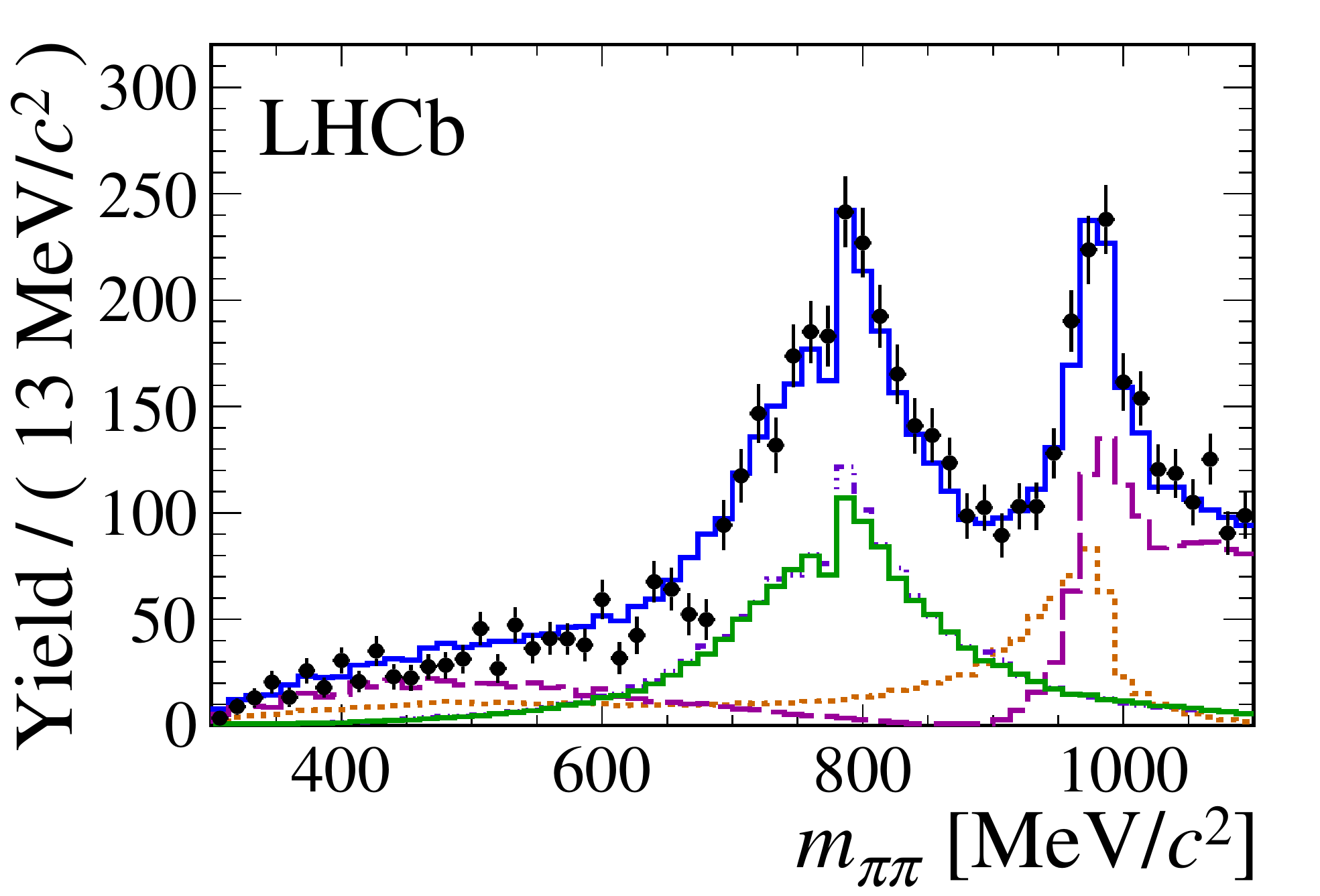}
\includegraphics[width=\ww]{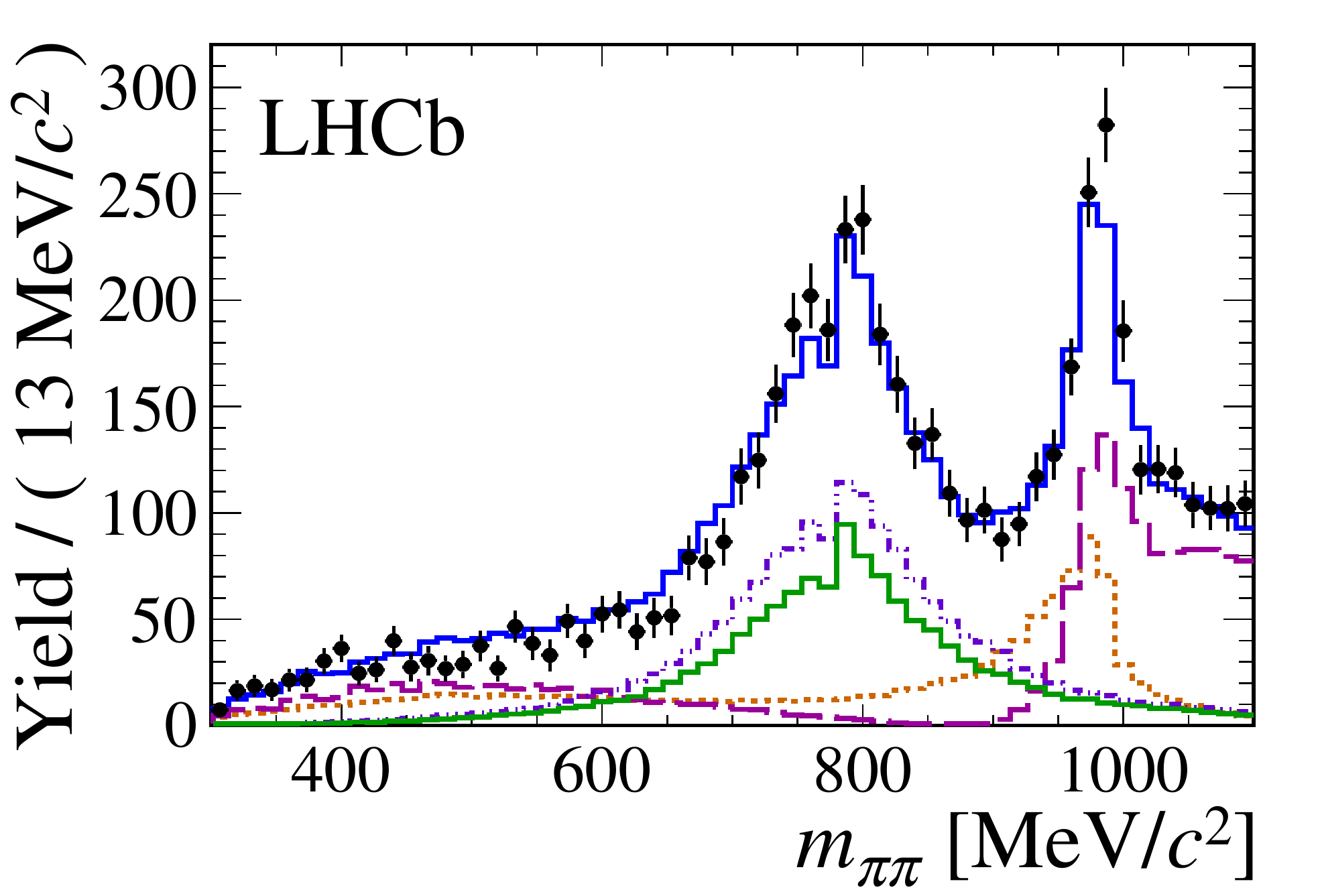}\\
\includegraphics[width=\ww]{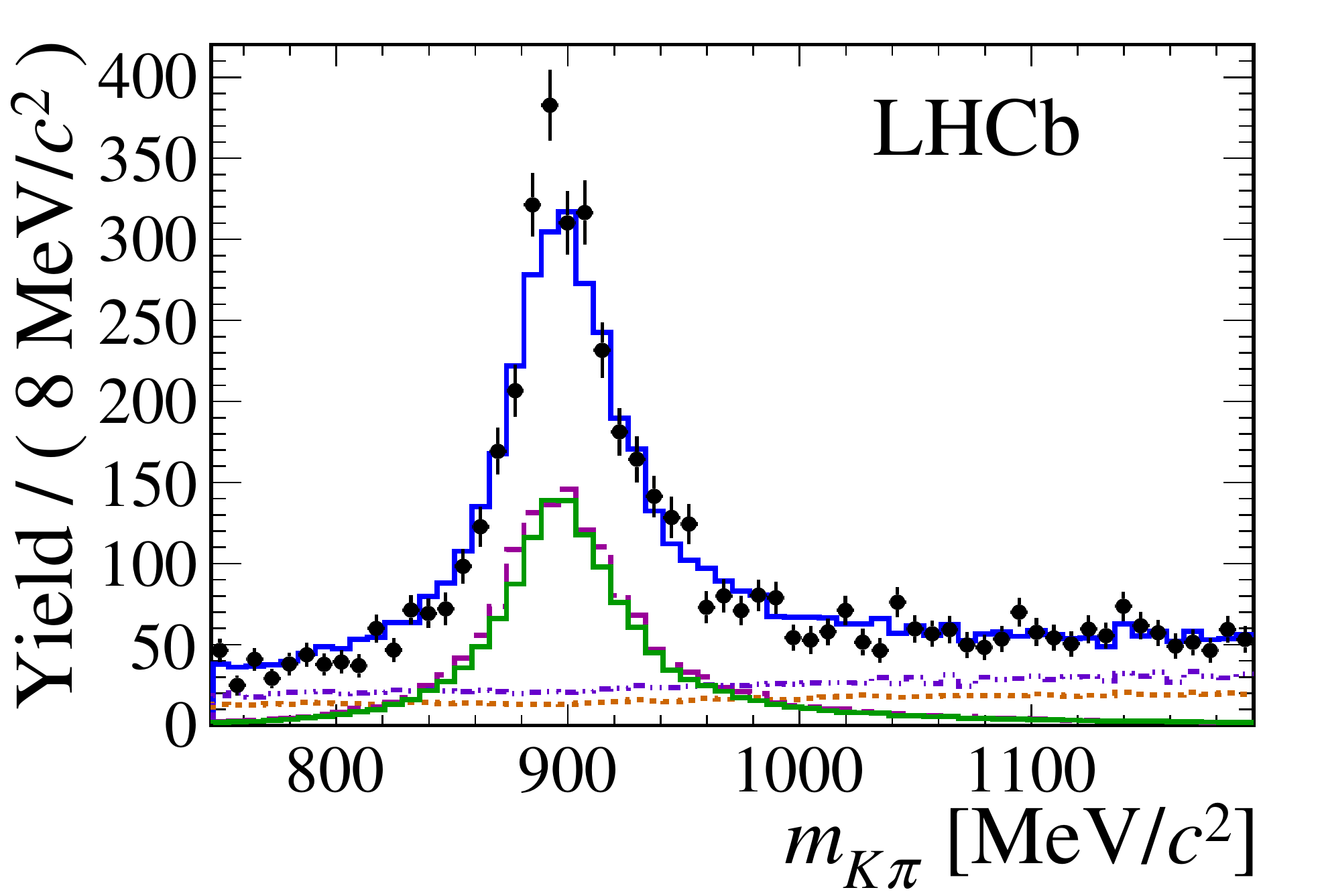}
\includegraphics[width=\ww]{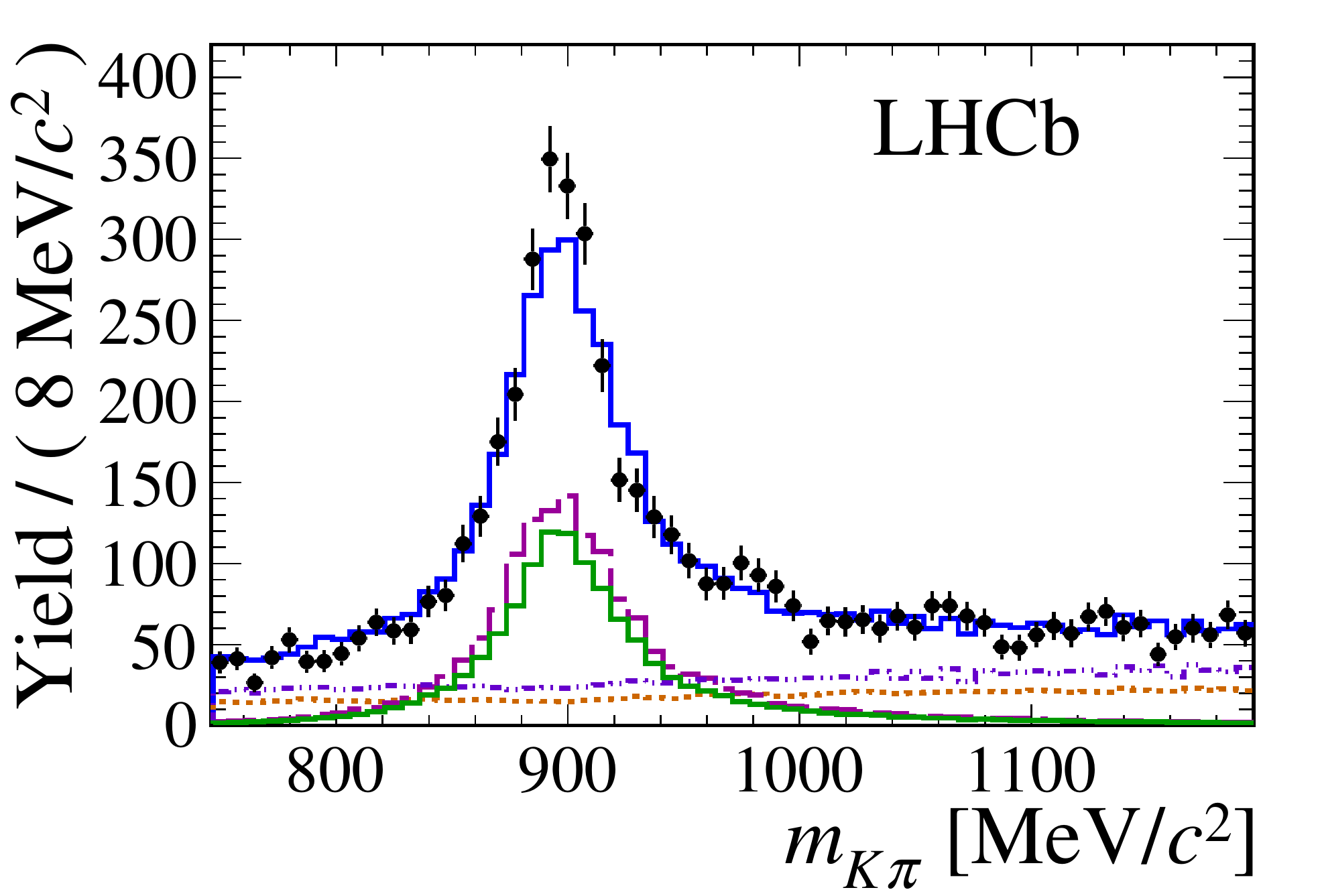}\\
\caption{\small{Projections of the amplitude fit to the (left) \Bd and (right) \Bdb data samples. The four trigger and data-taking year categories are aggregated in the figures. Data are shown by black points with uncertainties and the overall fit is represented by the solid blue line. The contributions of the partial waves sharing the same angular dependence are shown as ($VV$) solid green, ($VS$) dash-dotted violet, ($SV$)  dashed dark magenta and ($SS$) dotted orange lines. Direct \CP-violating effects are most visible in the projections of the $VV$ component over $\cos\theta_{K\pi}$ and $\cos\theta_{\pi\pi}$ and in the different oscillation frequency in $\phi$.}}
\label{fig:fitBBbar}
\end{centering}
\end{figure}

The one-dimensional projections of the maximum-likelihood fit to the \Bd and \Bdb weighted data samples are shown in \fig{fitBBbar}. The contribution of each partial wave is also shown. The fit results and their related observables, together with their statistical and total systematic uncertainties, anticipated from~\refsec{systs}, are reported in \tab{res}. The statistical uncertainties on all the reported quantities are evaluated using pseudoexperiments to properly account for possible nonlinear correlations among the parameters. The amplitude fit is repeated using subsets of the total data sample, employing only one of the trigger categories or data from one of the data-taking periods, yielding compatible results within statistical uncertainties. 

Using the nominal results and \eq{TPA} the following values of the TPA are found
\begin{align}
\mathcal{A}_{\text{T-\it{fake}}}^{\rhomeson\Kstar,1} &=\psm0.042\pz \pm 0.005\pz \pm 0.005,\pz \quad\quad\quad\quad
\mathcal{A}_{\text{T-\it{fake}}}^{\rhomeson\Kstar,2} &=-0.004\pz \pm 0.006\pz \pm 0.007,\pz \nonumber\\
\mathcal{A}_{\text{T-\it{fake}}}^{\omegaz\Kstar,1} &=\psm0.04\pzz \pm 0.04\pzz\pm 0.04,\pzz \quad\quad\quad\quad
\mathcal{A}_{\text{T-\it{fake}}}^{\omegaz\Kstar,2} &=-0.005\pz \pm 0.021\pz \pm 0.023,\pz
\nonumber
\end{align}

\begin{align}
\mathcal{A}_{\text{T-\it{true}}}^{\rhomeson\Kstar,1} &=-0.0210 \pm 0.0050 \pm 0.0022, \quad\quad\quad\quad
\mathcal{A}_{\text{T-\it{true}}}^{\rhomeson\Kstar,2} &=-0.003\pz \pm 0.006\pz \pm 0.005,\pz
\nonumber\\
\mathcal{A}_{\text{T-\it{true}}}^{\omegaz\Kstar,1} &=\psm0.022\pz \pm 0.043\pz \pm 0.016,\pz \quad\quad\quad\quad
\mathcal{A}_{\text{T-\it{true}}}^{\omegaz\Kstar,2} &=-0.014\pz \pm 0.021\pz \pm 0.017,\pz
\nonumber
\end{align}
where the first uncertainty is statistical and the second systematic. These results are compatible with SM expectations of TPAs below approximately $5\%$ for charmless {\decay{\Bd}{VV}} meson decays~\cite{Datta:2003mj}. Nevertheless, theoretical predictions of TPAs in exclusive decays are strongly affected by the knowledge of the nonfactorisable terms in the helicity amplitudes due to long-distance effects. The measurements reported above add valuable information in this regard.

\begin{table}[!ht]\centering
\caption{\small{Numerical fit results for the \CP averages and asymmetries in the (top) modulus and (bottom) phase differences of all the contributing amplitudes and among the $VV$ polarisation fractions. For the numbers in the table, the first and second uncertainties correspond to the statistical and total systematic, respectively. The total systematic uncertainty is obtained from the sum in quadrature of the individual sources detailed in \refsec{systs}, accounting for 100\% correlation of the common systematic uncertainties for \Bd and \Bdb.}}\label{tab:res}
\resizebox{0.81\textwidth}{!}{
\begin{tabular}{c|c|c}
\toprule
{Parameter} & {\cp~average, $\tilde{f}$} & {\cp~asymmetry, $\mathcal{A}$}\\
\midrule
$|A_{\rhomeson\Kstar}^{0}|^{2}  $  & $  \psm0.32\pzz \pm 0.04\pzz \pm 0.07\pzz $  & $  -0.75\pz \pm 0.07\pz \pm 0.17\pz$\\
$|A_{\rhomeson\Kstar}^{||}|^{2}  $  & $  \psm0.70\pzz \pm 0.04\pzz \pm 0.08\pzz $  & $  -0.049 \pm 0.053 \pm 0.019$\\
$|A_{\rhomeson\Kstar}^{\perp}|^{2}  $  & $  \psm0.67\pzz \pm 0.04\pzz \pm 0.07\pzz $  & $  -0.187 \pm 0.051 \pm 0.026$\\
$|A_{\omegaz\Kstar}^{0}|^{2}  $  & $  \psm0.019\pz \pm 0.010\pz \pm 0.012\pz $  & $  -0.6\pzz \pm 0.4\pzz \pm 0.4\pzz$\\
$|A_{\omegaz\Kstar}^{||}|^{2}  $  & $ \psm0.0050 \pm 0.0029 \pm 0.0031 $  & $  -0.30\pz \pm 0.54\pz \pm 0.28\pz$\\
$|A_{\omegaz\Kstar}^{\perp}|^{2}  $  & $ \psm0.0020 \pm 0.0019 \pm 0.0015 $  & $  -0.2\pzz \pm 0.9\pzz \pm 0.4\pzz$\\
$|A_{\omegaz (\kaon\pion)}|^{2}  $  & $  \psm0.026\pz \pm 0.011\pz \pm 0.025\pz $  & $  -0.47\pz \pm 0.33\pz \pm 0.45\pz$\\
$|A_{f_0(500)\Kstar}|^{2}  $  & $  \psm0.53\pzz \pm 0.05\pzz \pm 0.10\pzz$  & $  -0.06\pz \pm 0.09\pz \pm 0.04\pz$\\
$|A_{f_0(980)\Kstar}|^{2}  $  & $  \psm2.42\pzz \pm 0.13\pzz \pm 0.25\pzz $  & $  -0.022 \pm 0.052 \pm 0.023$\\
$|A_{f_0(1370)\Kstar}|^{2}  $  & $  \psm1.29\pzz \pm 0.09\pzz \pm 0.20\pzz $  & $  -0.09\pz \pm 0.07\pz \pm 0.04\pz$\\
$|A_{f_0(500)(\kaon\pion)}|^{2}  $  & $  \psm0.174\pz \pm 0.021\pz \pm 0.039\pz $  & $  \psm0.30\pz \pm 0.12\pz \pm 0.09\pz$\\
$|A_{f_0(980)(\kaon\pion)}|^{2}  $  & $  \psm1.18\pzz \pm 0.08\pzz \pm 0.07\pzz $  & $  -0.083 \pm 0.066 \pm 0.023$\\
$|A_{f_0(1370)(\kaon\pion)}|^{2}  $  & $  \psm0.139\pz \pm 0.028\pz \pm 0.039\pz $  & $  -0.48\pz \pm 0.17\pz \pm 0.15\pz$\\
\midrule
$f_{\rhomeson\Kstar}^{0}  $  & $  \psm0.164\pz \pm 0.015\pz \pm 0.022\pz  $  & $  -0.62\pz \pm 0.09\pz \pm 0.09\pz$\\
$f_{\rhomeson\Kstar}^{||}  $  & $  \psm0.435\pz \pm 0.016\pz \pm 0.042\pz  $  & $  \psm0.188 \pm 0.037 \pm 0.022$\\
$f_{\rhomeson\Kstar}^{\perp}  $  & $  \psm0.401\pz \pm 0.016\pz \pm 0.037\pz  $  & $  \psm0.050 \pm 0.039 \pm 0.015$\\
$f_{\omegaz\Kstar}^{0}  $  & $  \psm0.68\pzz \pm 0.17\pzz \pm 0.16\pzz  $  & $  -0.13\pz \pm 0.27\pz \pm 0.13\pz$\\
$f_{\omegaz\Kstar}^{||}  $  & $  \psm0.22\pzz \pm 0.14\pzz \pm 0.15\pzz  $  & $  \psm0.26\pz \pm 0.55\pz \pm 0.22\pz$\\
$f_{\omegaz\Kstar}^{\perp}  $  & $  \psm0.10\pzz \pm 0.09\pzz \pm 0.09\pzz  $  & $  \psm0.3\pzz \pm 0.8\pzz \pm 0.4\pzz$\\
\midrule
{Parameter} & {\CP average, $\frac{1}{2}(\delta_{\Bbar} + \delta_B)$} [rad] & {\CP difference, $\frac{1}{2}(\delta_{\Bbar}-\delta_B)$ [rad]}\\
\midrule
$\delta_{\rhomeson\Kstar}^{0}  $  & $  \psm1.57\pzz \pm 0.08\pzz \pm 0.18\pzz  $  & $  \psm0.12\pz \pm 0.08\pz \pm 0.04\pz$\\
$\delta_{\rhomeson\Kstar}^{||}  $  & $  \psm0.795\pz \pm 0.030\pz \pm 0.068\pz  $  & $  \psm0.014 \pm 0.030 \pm 0.026$\\
$\delta_{\rhomeson\Kstar}^{\perp}  $  & $  -2.365\pz \pm 0.032\pz \pm 0.054\pz  $  & $  \psm0.000 \pm 0.032 \pm 0.013$\\
$\delta_{\omegaz\Kstar}^{0}  $  & $  -0.86\pzz \pm 0.29\pzz \pm 0.71\pzz  $  & $  \psm0.03\pz \pm 0.29\pz \pm 0.16\pz$\\
$\delta_{\omegaz\Kstar}^{||}  $  & $  -1.83\pzz \pm 0.29\pzz \pm 0.32\pzz  $  & $  \psm0.59\pz \pm 0.29\pz \pm 0.07\pz$\\
$\delta_{\omegaz\Kstar}^{\perp}  $  & $  \psm1.6\pzz\pz \pm 0.4\pzz\pz \pm 0.6\pzz\pz  $  & $  -0.25\pz \pm 0.43\pz \pm 0.16\pz$\\
$\delta_{\omegaz (\kaon\pion)}  $  & $  -2.32\pzz \pm 0.22\pzz \pm 0.24\pzz  $  & $  -0.20\pz \pm 0.22\pz \pm 0.14\pz$\\
$\delta_{f_0(500)\Kstar}  $  & $  -2.28\pzz \pm 0.06\pzz \pm 0.22\pzz  $  & $  -0.00\pz \pm 0.06\pz \pm 0.05\pz$\\
$\delta_{f_0(980)\Kstar}  $  & $  \psm0.39\pzz \pm 0.04 \pzz \pm 0.07\pzz  $  & $  \psm0.018 \pm 0.038 \pm 0.022$\\
$\delta_{f_0(1370)\Kstar}  $  & $  -2.76\pzz \pm 0.05\pzz \pm 0.09\pzz  $  & $  \psm0.076 \pm 0.051 \pm 0.025$\\
$\delta_{f_0(500)(\kaon\pion)}  $  & $  -2.80\pzz \pm 0.09\pzz \pm 0.21\pzz  $  & $  -0.206 \pm 0.088 \pm 0.034$\\
$\delta_{f_0(980)(\kaon\pion)}  $  & $  -2.982\pz \pm 0.032\pz \pm 0.057\pz  $  & $  -0.027 \pm 0.032 \pm 0.013$\\
$\delta_{f_0(1370)(\kaon\pion)}  $  & $  \psm1.76\pzz \pm 0.10\pzz \pm 0.11\pzz  $  & $  -0.16\pz \pm 0.10\pz \pm 0.04\pz$\\
\midrule
$\delta_{\rhomeson\Kstar}^{||-\perp}  $  & $  \psm3.160\pz \pm 0.035\pz\pm 0.044\psm  $  & $  \psm0.014 \pm 0.035 \pm 0.026$\\
$\delta_{\rhomeson\Kstar}^{||-0}  $  & $  -0.77\pzz \pm 0.09\pzz\pm 0.06\pzz  $  & $  -0.109 \pm 0.085 \pm 0.034$\\
$\delta_{\rhomeson\Kstar}^{\perp-0}  $  & $  -3.93\pzz \pm 0.09\pzz\pm 0.07\pzz  $  & $  -0.123 \pm 0.085 \pm 0.035$\\
$\delta_{\omegaz\Kstar}^{||-\perp}  $  & $  -3.4\pzz\pz \pm 0.5\pzz\pz\pm 0.7\pzz\pz  $  & $  \psm0.84\pz \pm 0.52\pz \pm 0.16\pz$\\
$\delta_{\omegaz\Kstar}^{||-0}  $  & $  -1.0\pzz\pz \pm 0.4\pzz\pz\pm 0.6\pzz\pz  $  & $  \psm0.57\pz \pm 0.41\pz \pm 0.17\pz$\\
$\delta_{\omegaz\Kstar}^{\perp-0}  $  & $  \psm2.4\pzz\pz \pm 0.5\pzz\pz\pm 0.8\pzz\pz  $  & $  -0.28\pz \pm 0.51\pz \pm 0.24\pz$\\
\bottomrule
\end{tabular}}
\end{table}

\section{Systematic uncertainties}
\label{sec:systs}

Several sources of systematic uncertainty are considered. In some cases their impact on the measurements is evaluated by means of pseudoexperiments, which are simulated samples having the same size as the analysed data sample and generated from the PDF. 

\begin{description}
\item {\bf Uncertainties on the parameters of the mass propagators.}
To assess the effect of the uncertainty in the mass, width and radii of the (\pipic) and (\kpic) propagators, a pseudoexperiment is generated with the default values used in the nominal fit. This sample is fitted two hundred times using alternative values for these parameters generated according to their known uncertainties. The distribution of all the values obtained for each observable is fitted with a Gaussian function whose width is taken as the systematic uncertainty. 

\item {\bf Angular momentum barrier factors.}     
As introduced in \refsec{ampfit}, the angular barrier factors arising from the production of the vector meson candidates are neglected. However, \P-odd states and the $VS/SV$ decay channels are only allowed to be produced with relative orbital angular momentum $L = 1$, while the $VV$ \P-even transversity amplitudes both contain superposition of $L = 0 $ and $L = 2$ orbital angular momentum states. These other configurations are allowed and the largest difference between the nominal and alternative fit results is assigned as a systematic uncertainty. 

\item {\bf Background subtraction.}
To account for uncertainties in the background subtraction, the parameters of the Hypatia distributions are varied according to their uncertainties and the yield of \bskstkst misidentified events is varied by $\pm2\sigma$, along with the weights applied to cancel this background component.
The four-body invariant-mass fit is repeated two hundred times to obtain alternative sets of signal weights accounting for each of the two sets of variations introduced. These are propagated to the amplitude fit and a systematic uncertainty assigned as described in the first item.

\item {\bf Description of the kinematic acceptance.}
Normalisation weights are obtained from simulated samples of limited size. Their statistical uncertainty is considered by using in the amplitude fit two hundred sets of alternative weights generated according to their covariance matrix. 

\item {\bf Masses and angular resolution.}
In the nominal fit the resolution of the five observables is neglected.
The systematic uncertainty due to this approximation is evaluated with pseudoexperiments. An ensemble of four hundred pseudoexperiments is generated and fitted before and after being smeared according to the resolution determined from simulation. The bias produced in the amplitude results is used to asses this uncertainty.
 
\item {\bf Fit method.} 
A collection of eight hundred pseudoexperiments with the same number of candidates as observed in data is generated and fitted using the nominal PDF to evaluate biases induced by the fitting method. 

\item {\bf Pollution due to }\bm{{\decay{\Bd}{a_1(1260)^-\Kp}}} {\bf decays.} The same final state can also be produced by the {\decay{\Bd}{a_1(1260)^-\Kp}} decay followed by the {\decay{a_1(1260)^-}{\pipic\pim}} process.
They are strongly suppressed in the analysed data sample due to the selected range of the two-body invariant-mass pairs, but even a small pollution ($\sim 4\%$ relative amplitude with respect to the \brhokst~channel) may affect the results, due to the interference terms. Three sets of four hundred pseudoexperiments are generated with a pollution level compatible with data distributions. These three sets differ in the phase difference between the $a_1(1260)^-$ contribution and the reference amplitude, covering different interference patterns (0, 2$\pi$/3 and 4$\pi$/3). The maximum shift induced in the fit parameters is assigned as the corresponding systematic uncertainty. Other three-body decaying resonance contributions, such as {\decay{\Bd}{K_1(1270)^+\pim}}, are found to be fully rejected by the two-body invariant-mass requirements. 

\item {\bf{ Symmetrised (}}{\bm{$\pi\pi$}}{\bf{) contribution in the model.}}
The two same-charge pions in the final state may be exchanged and the PDF re-evaluated. This combination does not fulfil the invariant-mass requirements on both quasi-two-body systems but the interference between both configurations might give rise to some effect on the fit parameters, which is evaluated by generating four hundred pseudoexperiments and comparing the results of fitting with and without this contribution. 

\item {\bf Simulation corrections.}
Differences in the distributions of the \Bd  momentum, event multiplicity and the PID variables are observed between data and simulation and corrected for. Data is employed to obtain bidimensional efficiency maps, in bins of track pseudorapidity and momentum, for each year of data taking and magnet polarity. These maps are used to evaluate the PID track efficiency and to assign to each candidate a global PID efficiency weight. Furthermore, a second iterative method \cite{GarciaPardinas:2630181}, is used to weight the simulated events and improve the description of the track multiplicity and \Bd momentum distributions. The final fit results are obtained with the weights from the last iteration, and their difference with respect to those obtained using the weights from the previous to last iteration is assigned as the systematic uncertainty.
\end{description}

The resulting systematic uncertainties are reported in \tabs{systs1}{systs2} in \app{syststables}. The pollution due to {\decay{\Bd}{a_1(1260)^-\Kp}} decays represents the largest source of systematic uncertainty for the parameters related to the $VV$ waves, while the uncertainty on the parameters used in the mass propagators and the  resolution effects dominate the systematic uncertainties of the parameters related to the various $S$-waves. 

\section{Summary and conclusions}
\label{sec:conclusions}
The first full amplitude analysis of {\decay{\Bd}{(\pipic)(\kpic)}} decays in the two-body invariant mass windows of ${300 < m(\pipic)<1100}$\mevcc and ${750 < m(\kpic)<1200}$\mevcc is presented. The fit model is built using the isobar approach and accounts for 10 decay channels leading to a total of 14 interfering amplitudes. A remarkably small longitudinal polarisation fraction and a significant direct \CP asymmetry are measured for the \bdrhokst mode, hinting at a relevant contribution from the colour-allowed electroweak-penguin amplitude,  
\begin{equation*}
\tilde{f}^0_{\rhomeson\Kstar} = 0.164 \pm 0.015 \pm 0.022\quad \text{and}\quad{\mathcal{A}^0_{\rhomeson\Kstar} = -0.62 \pm 0.09 \pm 0.09}\, ,
\end{equation*}
where the first uncertainty is statistical and the second, systematic. The significance of the \CP asymmetry is obtained by dividing the value of the asymmetry by the sum in quadrature of the statistical and systematic uncertainties and is found to be in excess of 5 standard deviations. This is the first significant observation of \CP asymmetry in angular distributions of {\decay{\Bd}{VV}} decays.
A determination of the equivalent parameters for the \bomkst mode is also made, resulting in
\begin{equation*}
    \Tilde{f}^0_{\omegaz\Kstar} = 0.68 \pm 0.17 \pm 0.16 \quad \text{and}\quad {\mathcal{A}^0_{\omegaz\Kstar} = -0.13 \pm 0.27 \pm 0.13}\, .
\end{equation*}
The phase differences between the perpendicular and parallel polarisation, $\delta^{||-\perp}_{\rhomeson\Kstar}$, are found to be very close to $\pi$ and $0$, for the \CP averaged and \CP difference values, respectively. These are in good agreement with theoretical predictions computed in both QCDF and pQCD frameworks. \tab{thpred} shows a comparison among the results obtained in this analysis and the most recent predictions in these two theoretical approaches.
\begin{table}
\caption{\small{Comparison of theoretical predictions for the \bdrhokst mode with the results obtained from this analysis. It should be noted that the theoretical predictions involving the \CP averaged value of $\delta^\perp_{\rhomeson\Kstar}$ have been shifted by $\pi$ on account of the different phase conventions used in the theoretical and experimental works.}}
\label{tab:thpred}
\begin{center}
\resizebox{0.98\textwidth}{!}{
\begin{tabular}{crccc}
\toprule
\multicolumn{2}{c}{Observable}         & QCDF~\cite{Beneke:2006hg} & pQCD~\cite{Zou:2015iwa} & This work \\
\midrule
\\[-0.15em]
\parbox[t]{3mm}{\multirow{2}{*}{\rotatebox[origin=c]{90}{$f^0_{\rhomeson\Kstar}$}}}&\CP average & $\psm0.22^{+0.03 +0.53}_{-0.03-0.14}$ & $0.65^{+0.03+0.03}_{-0.03-0.04}$ & $\psm0.164 \pm 0.015 \pm 0.022$ \\
\\[-0.15em]
&\CP asymmetry & $-0.30^{+0.11+0.61}_{-0.11-0.49}$ & $0.0364^{+0.0120}_{-0.0107}$ & $-0.62\phantom{2} \pm 0.09\phantom{0} \pm 0.09\phantom{0}$\\
\\[-0.05em]
\midrule
\\[-0.05em]
\parbox[t]{3mm}{\multirow{2}{*}{\rotatebox[origin=c]{90}{$f^\perp_{\rhomeson\Kstar}$}}}&\CP average & $0.39^{+0.02+0.27}_{-0.02-0.07}$ & $\psm0.169\pz^{+0.027}_{-0.018}$ & $\psm0.401 \pm 0.016 \pm 0.037$ \\
\\[-0.15em]
&\CP asymmetry & $-$ & $-0.0771^{+0.0197}_{-0.0186}$ & $\psm0.050 \pm 0.039 \pm 0.015$ \\
\\[-0.05em]
\midrule
\\[-0.05em]
\parbox[t]{3mm}{\multirow{2}{*}{\rotatebox[origin=c]{90}{$\delta^{||-0}_{\rhomeson\Kstar}$ }}}& \CP average [rad] & $-0.7\pz^{+0.1+1.1}_{-0.1-0.8}$ & $-1.61\pz^{+0.02}_{-3.06}$ & $-0.77\phantom{0} \pm 0.09\phantom{0} \pm 0.06\phantom{0}$ \\  
\\[-0.15em]
&\CP difference [rad] & $\psm0.30^{+0.09+0.38}_{-0.09-0.33}$ & $-0.001^{+0.017}_{-0.018}$ & $-0.109 \pm 0.085 \pm 0.034$ \\
\\[-0.05em]
\midrule
\\[-0.05em]
\parbox[t]{3mm}{\multirow{2}{*}{\rotatebox[origin=c]{90}{$\delta^{||-\perp}_{\rhomeson\Kstar}$ }}}& \CP average [rad] & $\equiv\pi$ & $ \psm3.15\pz^{+0.02}_{-4.30}$ & $\psm3.160 \pm 0.035 \pm 0.044$ \\
\\[-0.15em]
&\CP difference [rad] & $\equiv0$ & $-0.003^{+0.025}_{-0.024}$ & $\psm0.014 \pm 0.035 \pm 0.026$\\
\\[-0.15em]
\bottomrule
\end{tabular}
}
\end{center}
\end{table}



\section*{Acknowledgements}
%
%
\noindent We express our gratitude to our colleagues in the CERN accelerator departments for the excellent performance of the LHC. We thank the technical and administrative staff at the LHCb institutes. We acknowledge support from CERN and from the national agencies:
CAPES, CNPq, FAPERJ and FINEP (Brazil); 
MOST and NSFC (China); 
CNRS/IN2P3 (France); 
BMBF, DFG and MPG (Germany); 
INFN (Italy); 
NWO (Netherlands); 
MNiSW and NCN (Poland); 
MEN/IFA (Romania); 
MSHE (Russia); 
MinECo (Spain); 
SNSF and SER (Switzerland); 
NASU (Ukraine); 
STFC (United Kingdom); 
NSF (USA).
We acknowledge the computing resources that are provided by CERN, IN2P3
(France), KIT and DESY (Germany), INFN (Italy), SURF (Netherlands),
PIC (Spain), GridPP (United Kingdom), RRCKI and Yandex
LLC (Russia), CSCS (Switzerland), IFIN-HH (Romania), CBPF (Brazil),
PL-GRID (Poland) and OSC (USA).
We are indebted to the communities behind the multiple open-source
software packages on which we depend.
Individual groups or members have received support from
AvH Foundation (Germany);
EPLANET, Marie Sk\l{}odowska-Curie Actions and ERC (European Union);
ANR, Labex P2IO and OCEVU, and R\'{e}gion Auvergne-Rh\^{o}ne-Alpes (France);
Key Research Program of Frontier Sciences of CAS, CAS PIFI, and the Thousand Talents Program (China);
RFBR, RSF and Yandex LLC (Russia);
GVA, XuntaGal and GENCAT (Spain);
the Royal Society
and the Leverhulme Trust (United Kingdom);
Laboratory Directed Research and Development program of LANL (USA).


\section*{Appendices}

\appendix

\section{Legend}
\label{app:legend}

\begin{figure}[ht!]
\begin{centering}
\includegraphics[width=0.6\textwidth]{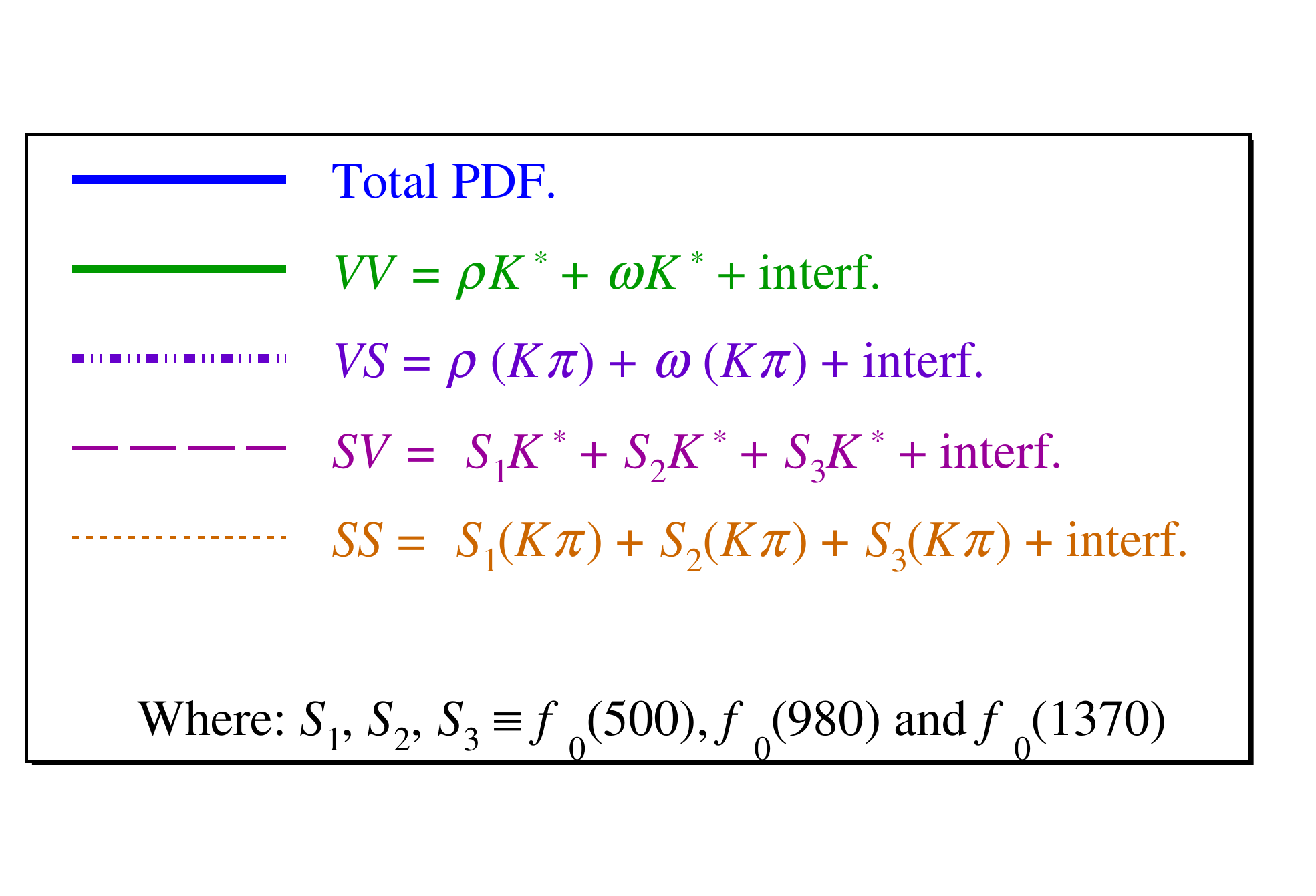}
\caption{\small Legend for the plots. The partial waves sharing the same angular dependence are represented as ($VV$) solid green, ($VS$) dash-dotted violet, ($SV$)  dashed dark magenta and ($SS$) dotted orange lines. The overall fit is shown by a solid blue line.}
\label{fig:legend}
\end{centering}
\end{figure}

\section{Breakdown of the systematic uncertainties}
\label{app:syststables}

In \tabs{systs1}{systs2} the break-up of systematic uncertainty contributions for the reported observables is shown.

\begin{sidewaystable}[t]
    \caption{\small Table (I) of the systematic uncertainties. The abbreviations $S1, S2$ and $S3$ stand for $f_0(500), f_0(980)$ and $f_0(1370)$, respectively. Negligible values are represented by a dash ($-$).}
    \label{tab:systs1}
    \centering
    \resizebox{\textwidth}{!}{

    \begin{tabular}{clS[table-format=1.5]S[table-format=1.5]S[table-format=1.5]S[table-format=1.5]S[table-format=1.5]S[table-format=1.5]S[table-format=1.5]S[table-format=1.5]S[table-format=1.5]S[table-format=1.5]}
    \toprule
 & {Systematic uncertainty} & {$|A_{\rhomeson\Kstar}^{0}|^{2}$} & {$|A_{\rhomeson\Kstar}^{||}|^{2}$} & {$|A_{\rhomeson\Kstar}^{\perp}|^{2}$} & {$|A_{\omegaz\Kstar}^{0}|^{2}$} & {$|A_{\omegaz\Kstar}^{||}|^{2}$} & {$|A_{\omegaz\Kstar}^{\perp}|^{2}$} & {$|A_{\omegaz (\kpin)}|^{2}$} & {$|A_{S1\Kstar}|^{2}$} & {$|A_{S2\Kstar}|^{2}$} & {$|A_{S3\Kstar}|^{2}$} \\

 \midrule 
 
\parbox[t]{3mm}{\multirow{5}{*}{\rotatebox[origin=c]{90}{\CP averages}}} & Centrifugal barrier factors  & {$-$\phantom{.}}  & {$-$\phantom{.}}  &  {$-$\phantom{.}}  &  {$-$\phantom{.}}  &  0.0001  &  {$-$\phantom{.}}  &  0.001  &  0.01  &  0.01  &  0.04\\
& Hypatia parameters & {$-$\phantom{.}}  &  {$-$\phantom{.}}  &  {$-$\phantom{.}}  &  {$-$\phantom{.}}  &  {$-$\phantom{.}}  &  {$-$\phantom{.}}  &  {$-$\phantom{.}}  &  {$-$\phantom{.}}  &  {$-$\phantom{.}}  &  {$-$\phantom{.}}\\
& \bskstkst bkg. &  0.01  &  0.01  &  0.01  &  0.001  &  0.0004  &  0.0002  &  0.001  &  0.01  &  0.02  &  0.01\\
& Simulation sample size & 0.01  &  0.01  &  0.01  &  0.002  &  0.0007  &  0.0003  &  0.005  &  0.02  &  0.06  &  0.04\\
&Data-Simulation corrections & {$-$\phantom{.}}  &  {$-$\phantom{.}}  &  {$-$\phantom{.}}  &  {$-$\phantom{.}}  &  0.0002  &  {$-$\phantom{.}}  &  {$-$\phantom{.}}  &  {$-$\phantom{.}}  &  {$-$\phantom{.}}  &  {$-$\phantom{.}}\\

      \midrule 
  
\parbox[t]{3mm}{\multirow{5}{*}{\rotatebox[origin=c]{90}{\CP asym.}}} &Centrifugal barrier factors    & {$-$\phantom{.}}  &  {$-$\phantom{.}}  &  0.004  &  {$-$\phantom{.}}  &  {$-$\phantom{.}}  &  {$-$\phantom{.}}  &  0.01  &  {$-$\phantom{.}}  &  0.003  &  0.01\\
& Hypatia parameters & {$-$\phantom{.}}  &  0.002  &  0.002  &  {$-$\phantom{.}}  &  0.01  &  {$-$\phantom{.}}  &  0.01  &  {$-$\phantom{.}}  &  0.002  &  {$-$\phantom{.}}\\
& \bskstkst bkg. & 0.03  &  0.011  &  0.013  &  {$-$\phantom{.}}  &  0.13  &  0.1  &  0.01  &  0.02  &  0.005  &  0.01\\
& Simulation sample size  & 0.02  &  0.014  &  0.011  &  0.1  &  0.17  &  0.4  &  0.14  &  0.04  &  0.022  &  0.03\\
&Data-Simulation corrections  & {$-$\phantom{.}}  &  0.001  &  {$-$\phantom{.}}  &  {$-$\phantom{.}}  &  0.01  &  {$-$\phantom{.}}  &  0.01  &  {$-$\phantom{.}}  &  {$-$\phantom{.}}  &  {$-$\phantom{.}}\\

      \midrule 
\parbox[t]{5mm}{\multirow{5}{*}{\rotatebox[origin=c]{90}{\parbox{1.5cm}{Common\\ (\Bd,\Bdb)}}}}&  Mass propagators parameters  & 0.01 & 0.033 & 0.040 & 0.002 & 0.0003 & 0.0001 & 0.002 & 0.07 & 0.170 & 0.12\\
& Masses and angles resolution  & 0.01 & 0.023 & 0.040 & 0.010 & 0.0028 & 0.0010 & 0.024 & 0.03 & 0.050 & 0.10\\
& Fit method & 0.01 & 0.007 & 0.007 & 0.004 & 0.0005 & 0.0010 & 0.001 & 0.01 & 0.029 & {$-$\phantom{.}} \\
&  $a_1(1260)$ pollution & 0.06 & 0.070 & 0.019 & 0.003 & 0.0005 & 0.0002 & 0.003 & 0.05 & 0.130 & 0.10\\
&  Symmetrised (\pipin) PDF &0.04 & 0.030 & 0.021 & {$-$\phantom{.}} & 0.0008 & 0.0003 & 0.004 & 0.03 & 0.080 & 0.06 \\

\toprule

& {Systematic uncertainty }&{ $|A_{S1(\kpin)}|^{2}$ }&{ $|A_{S2(\kpin)}|^{2}$ }&{ $|A_{S3(\kpin)}|^{2}$ }&{ $\delta_{\rhomeson\Kstar}^{0}$ }&{ $\delta_{\rhomeson\Kstar}^{||}$ }&{ $\delta_{\rhomeson\Kstar}^{\perp}$ }&{ $\delta_{\omegaz\Kstar}^{0}$ }&{ $\delta_{\omegaz\Kstar}^{||}$ }&{ $\delta_{\omegaz\Kstar}^{\perp}$ }&{ $\delta_{\omegaz(\kpin)}$} \\
 
 \midrule 
 
\parbox[t]{3mm}{\multirow{5}{*}{\rotatebox[origin=c]{90}{\CP averages}}} & Centrifugal barrier factors   & 0.003  &  0.02  &  0.003  &  {$-$\phantom{.}}  &  0.001  &  0.002  &  0.03  &  0.01  &  {$-$\phantom{.}}  &  0.01\\
& Hypatia parameters & 0.001  &  0.01  &  0.001  &  {$-$\phantom{.}}  &  0.001  &  0.002  &  0.01  &  0.01  &  {$-$\phantom{.}}  &  {$-$\phantom{.}}\\
& \bskstkst bkg. & 0.008  &  0.01  &  0.004  &  0.02  &  0.018  &  0.007  &  0.04  &  0.02  &  0.1  &  0.01\\
& Simulation sample size  & 0.006  &  0.03  &  0.007  &  0.02  &  0.009  &  0.008  &  0.15  &  0.07  &  0.1  &  0.10\\
&Data-Simulation corrections  & {$-$\phantom{.}}  &  {$-$\phantom{.}}  &  0.001  &  {$-$\phantom{.}}  &  0.001  &  {$-$\phantom{.}}  &  {$-$\phantom{.}}  &  {$-$\phantom{.}}  &  {$-$\phantom{.}}  &  {$-$\phantom{.}}\\

      \midrule 
  
\parbox[t]{3mm}{\multirow{5}{*}{\rotatebox[origin=c]{90}{\CP asym.}}} &Centrifugal barrier factors  & {$-$\phantom{.}}  &  0.010  &  0.02  &  {$-$\phantom{.}}  &  0.004  &  0.001  &  0.02  &  0.01  &  0.03  &  0.02\\
& Hypatia parameters & 0.01  &  0.004  &  0.01  &  {$-$\phantom{.}}  &  0.001  &  0.001  &  0.01  &  0.01  &  0.01  &  {$-$\phantom{.}} \\
& \bskstkst bkg. &  0.05  &  0.007  &  0.03  &  0.03  &  0.024  &  0.009  &  0.05  &  0.02  &  0.06  &  0.02\\
& Simulation sample size  &  0.04  &  0.020  &  0.06  &  0.02  &  0.009  &  0.009  &  0.15  &  0.07  &  0.15  &  0.13\\
&Data-Simulation corrections  & {$-$\phantom{.}}  &  0.001  &  {$-$\phantom{.}}  &  {$-$\phantom{.}}  &  {$-$\phantom{.}}  &  {$-$\phantom{.}}  &  {$-$\phantom{.}}  &  0.01  &  0.01  &  {$-$\phantom{.}} \\

      \midrule 
\parbox[t]{5mm}{\multirow{5}{*}{\rotatebox[origin=c]{90}{\parbox{1.5cm}{Common\\ (\Bd,\Bdb)}}}}&  Mass propagators parameters  & 0.012 & 0.027 & 0.024 & 0.03 & 0.009 & 0.008 & 0.04 & 0.05 & 0.09 & 0.04 \\
& Masses and angles resolution  & 0.010 & 0.026 & 0.011 & 0.03 & 0.020 & 0.017 & 0.30 & 0.30 & 0.50 & 0.17\\
& Fit method  & 0.003 & 0.021 & 0.005 & {$-$\phantom{.}} & 0.001 & 0.001 & 0.03 & 0.05 & 0.04 & 0.01\\
&  $a_1(1260)$ pollution  & 0.018 & 0.040 & 0.019 & 0.17 & 0.060 & 0.050 & 0.60 & 0.06 & 0.05 & 0.12\\
&  Symmetrised (\pipin) PDF  & 0.029 & 0.025 & 0.019 & 0.02 & 0.010 & 0.012 & {$-$\phantom{.}} & 0.04 & 0.30 & 0.05 \\

\bottomrule
    \end{tabular}}
\end{sidewaystable}

\begin{sidewaystable}[t]

\caption{\small Table (II) of the systematic uncertainties. The abbreviations $S1, S2$ and $S3$ stand for $f_0(500), f_0(980)$ and $f_0(1370)$, respectively. Negligible values are represented by a dash ($-$).}
    \label{tab:systs2}
    \centering
    \resizebox{\textwidth}{!}{
    \begin{tabular}{clS[table-format=1.5]S[table-format=1.5]S[table-format=1.5]S[table-format=1.5]S[table-format=1.5]S[table-format=1.5]S[table-format=1.5]S[table-format=1.5]S[table-format=1.5]S[table-format=1.5]S[table-format=1.5]}
    \toprule
& {Systematic uncertainty}&{ $\delta_{S1\Kstar}$ }&{ $\delta_{S2\Kstar}$ }&{ $\delta_{S3\Kstar}$ }&{ $\delta_{S1(\kpin)}$ }&{ $\delta_{S2(\kpin)}$ }&{ $\delta_{S3(\kpin)}$ }&{ $f^{0}_{\rho\Kstar}$ }&{ $f^{||}_{\rho\Kstar}$ }&{ $f^{\perp}_{\rho\Kstar}$ }&{ $f^{0}_{\omegaz\Kstar}$} & {$f^{||}_{\omegaz\Kstar}$}  \\

  \midrule   
 
 \parbox[t]{3mm}{\multirow{5}{*}{\rotatebox[origin=c]{90}{\CP averages}}} & Centrifugal barrier factors  &  0.01  &  {$-$\phantom{.}}  &  0.01  &  0.01  &  0.001  &  0.02  &  0.001  &  0.001  &  0.002  &  {$-$\phantom{.}}  &  {$-$\phantom{.}} \\
& Hypatia parameters & {$-$\phantom{.}}  &  {$-$\phantom{.}}  &  {$-$\phantom{.}}  &  {$-$\phantom{.}}  &  0.001  &  0.01  &  0.001  &  0.001  &  0.001  &  {$-$\phantom{.}}  &  {$-$\phantom{.}} \\
& \bskstkst bkg. &0.05  &  {$-$\phantom{.}}  &  0.01  &  0.02 &  0.002  &  0.01  &  0.005  &  0.003  &  0.005  &  0.02  &  0.02\\
& Simulation sample size  & 0.02  &  0.01  &  0.02  &  0.02  &  0.009  &  0.03  &  0.004  &  0.004  &  0.004  &  0.06  &  0.05\\
&Data-Simulation corrections  & {$-$\phantom{.}}  &  {$-$\phantom{.}}  &  {$-$\phantom{.}}  &  {$-$\phantom{.}}  &  0.001  &  {$-$\phantom{.}}  &  {$-$\phantom{.}}  &  {$-$\phantom{.}}  &  {$-$\phantom{.}}  &  0.01  &  {$-$\phantom{.}} \\

 \midrule   
 
 \parbox[t]{3mm}{\multirow{5}{*}{\rotatebox[origin=c]{90}{\CP asym.}}} & Centrifugal barrier factors  &0.01  &  0.001  &  0.001  &  0.004  &  0.003  &  0.02  &  {$-$\phantom{.}}  &  0.001  &  0.002  &  0.01  &  0.01\\
& Hypatia parameters &{$-$\phantom{.}}  &  0.002  &  0.002  &  0.004  &  0.001  &  0.01  &  {$-$\phantom{.}}  &  0.003  &  0.002  &  0.01  &  0.01\\
& \bskstkst bkg. & 0.04  &  0.005  &  0.011  &  0.023  &  0.002  &  0.01  &  0.03  &  0.007  &  0.011  &  0.03  &  0.06\\
& Simulation sample size  &0.03  &  0.022  &  0.022  &  0.025  &  0.012  &  0.03  &  0.02  &  0.010  &  0.009  &  0.12  &  0.14\\
&Data-Simulation corrections  & {$-$\phantom{.}}  &  0.001  &  {$-$\phantom{.}}  &  0.003  &  {$-$\phantom{.}}  &  {$-$\phantom{.}}  &  {$-$\phantom{.}}  &  0.001  &  0.001  &  {$-$\phantom{.}}  &  0.01\\

      \midrule 
\parbox[t]{5mm}{\multirow{5}{*}{\rotatebox[origin=c]{90}{\parbox{1.5cm}{Common\\ (\Bd,\Bdb)}}}}&  Mass propagators parameters  &0.19 & 0.031 & 0.070 & 0.200 & 0.018 & 0.06 & 0.011 & 0.005 & 0.006 & 0.01 & 0.01 \\
& Masses and angles resolution  &0.02 & 0.027 & 0.017 & 0.026 & 0.026 & 0.05 & 0.010 & 0.016 & 0.018 & 0.14 & 0.12 \\
& Fit method  & {$-$\phantom{.}} & 0.004 & 0.001 & 0.002 & 0.001 & {$-$\phantom{.}} & 0.003 & 0.001 & 0.002 & 0.01 & 0.05 \\
&  $a_1(1260)$ pollution  & 0.09 & 0.040 & 0.040 & 0.040 & 0.050 & 0.04 & 0.015 & 0.040 & 0.031 & 0.02 & 0.01 \\
&  Symmetrised (\pipin) PDF  &0.03 & 0.029 & 0.022 & 0.035 & 0.006 & 0.05 & 0.004 & {$-$\phantom{.}} & 0.004 & 0.04 & 0.05 \\

    \toprule
& {Systematic uncertainty}&{ $f^{\perp}_{\omegaz\Kstar}$ }&{  $\delta_{\rhomeson\Kstar}^{||-\perp}$ }&{ $\delta_{\rhomeson\Kstar}^{||-0}$ }&{ $\delta_{\rhomeson\Kstar}^{\perp-0}$ }&{ $\delta_{\omegaz\Kstar}^{||-\perp}$ }&{ $\delta_{\omegaz\Kstar}^{||-0}$ }&{ $\delta_{\omegaz\Kstar}^{\perp-0}$ }&{ $\mathcal{A}_{\text{T}}^{\rho\Kstar,1}$ }&{ $\mathcal{A}_{\text{T}}^{\rho\Kstar,2}$ }&{ $\mathcal{A}_{\text{T}}^{\omegaz\Kstar,1}$ }&{ $\mathcal{A}_{\text{T}}^{\omegaz\Kstar,2}$} \\

  \midrule   
 
 \parbox[t]{3mm}{\multirow{5}{*}{\rotatebox[origin=c]{90}{\CP averages}}} & Centrifugal barrier factors  & {$-$\phantom{.}}  &  0.001  &  {$-$\phantom{.}}  &  {$-$\phantom{.}}  &  {$-$\phantom{.}}  &  {$-$\phantom{.}}  &  {$-$\phantom{.}}  &  0.0002  &  {$-$\phantom{.}}  &  0.001  &  0.001\\
& Hypatia parameters & {$-$\phantom{.}}  &  0.001  &  {$-$\phantom{.}}  &  {$-$\phantom{.}}  &  {$-$\phantom{.}}  &  {$-$\phantom{.}}  &  {$-$\phantom{.}}  &  0.0002  &  {$-$\phantom{.}}  &  0.001  &  0.001\\
& \bskstkst bkg.  & 0.01  &  0.018  &  0.02  &  0.02  &  0.1  &  {$-$\phantom{.}}  &  0.1  &  0.0017  &  0.002  &  0.004  &  0.002\\
& Simulation sample size   &0.03  &  0.009  &  0.02  &  0.02  &  0.2  &  0.2  &  0.2  &  0.0013  &  0.002  &  0.012  &  0.012\\
&Data-Simulation corrections  & {$-$\phantom{.}}  &  0.001  &  {$-$\phantom{.}}  &  {$-$\phantom{.}}  &  {$-$\phantom{.}}  &  {$-$\phantom{.}}  &  {$-$\phantom{.}}  &  {$-$\phantom{.}}  &  {$-$\phantom{.}}  &  {$-$\phantom{.}}  &  {$-$\phantom{.}}\\

 \midrule   
 
 \parbox[t]{3mm}{\multirow{5}{*}{\rotatebox[origin=c]{90}{\CP asym.}}} & Centrifugal barrier factors   &{$-$\phantom{.}}  &  0.004  &  0.007  &  0.004  &  0.03  &  0.02  &  0.04  &  0.0003  &  0.001  &  0.001  &  0.001\\
& Hypatia parameters &0.1  &  0.001  &  0.002  &  0.002  &  0.02  &  0.01  &  0.02  &  0.0001  &  {$-$\phantom{.}}  &  0.001  &  0.001\\
& \bskstkst bkg. & 0.2  &  0.024  &  0.020  &  0.026  &  0.06  &  0.04  &  0.13  &  0.0017  &  0.004  &  0.005  &  0.003\\
& Simulation sample size   &0.1  &  0.011  &  0.027  &  0.023  &  0.14  &  0.17  &  0.20  &  0.0013  &  0.002  &  0.015  &  0.017\\
&Data-Simulation corrections  &{$-$\phantom{.}}  &  {$-$\phantom{.}}  &  0.002  &  0.002  &  0.02  &  0.01  &  0.01  &  {$-$\phantom{.}} &  {$-$\phantom{.}}  &  0.001  &  {$-$\phantom{.}}\\

      \midrule 
\parbox[t]{5mm}{\multirow{5}{*}{\rotatebox[origin=c]{90}{\parbox{1.5cm}{Common\\ (\Bd,\Bdb)}}}}&  Mass propagators parameters  & {$-$\phantom{.}} &0.004 & 0.028 & 0.024 & 0.07 & 0.06 & 0.09 & 0.0006 & 0.001 & 0.002 & {$-$\phantom{.}}\\
& Masses and angles resolution  & 0.08 & 0.031 & 0.029 & 0.040 & 0.60 & 0.40 & 0.60 & 0.0020 & 0.005 & 0.026 & 0.019\\
& Fit method  & 0.03 & 0.003 & 0.005 & 0.004 & 0.02 & 0.02 & 0.03 & 0.0001 & {$-$\phantom{.}} & 0.005 & 0.001\\
&  $a_1(1260)$ pollution  & 0.01& 0.024 & 0.035 & 0.032 & 0.24 & 0.32 & 0.40 & 0.0040 & 0.004 & 0.012 & 0.001\\
&  Symmetrised (\pipin) PDF   &0.03 & 0.005 & 0.001 & 0.001 & 0.35 & 0.02 & 0.29 & 0.0007 & 0.001 & 0.018 & 0.003\\

\bottomrule
    \end{tabular}}

\end{sidewaystable}

\newpage
\section{Phase-space density and two-body invariant-mass propagators}
\label{app:massprop}
\begin{table}[t]
\caption{Central values of the mass-propagator parameters and their uncertainties, used to estimate the corresponding systematic uncertainties. The values of the parameters used to describe the $f_0(500)$ and $f_0(1370)$ resonances were taken from Ref. \cite{Aaij:2014emv} and the rest, from Ref. \cite{PDG2018}.}
\label{tab:systprops}
\begin{center}
\begin{tabular}{l c }
\toprule
Parameter & Value\\
\midrule
$m_\rho$ [\mevcc] &775.26 $\pm$ 0.25\pzz		\\
$\Gamma_\rho$ [\mevcc] & \pz147.8 $\pm$ 0.9\pzz\pz	\\	
$r_{0_\rho}$[(\mevcc)$^{-1}$]&0.0053 $\pm$ 0.0008	\\
$m_{K^*}$ [\mevcc] &895.55	$\pm$ 0.20	\pzz		  \\     
$\Gamma_{K^*}$ [\mevcc] &\pzz47.3 $\pm$ 0.5\pzz\pz		\\
$r_{0_{K^*}}$[(\mevcc)$^{-1}$] &0.0030 $\pm$ 0.0005	\\
$m_\omega$ [\mevcc] &782.65 $\pm$ 0.12\pzz	\\	
$\Gamma_\omega$ [\mevcc] &\pzz8.49 $\pm$ 0.08\pzz\\		
$r_{0_\omega}$[(\mevcc)$^{-1}$]&0.0030 $\pm$ 0.0005   	\\
$m_{f_0(500)}$ [\mevcc] &\pzz475 $\pm$ 32\pzz\pz 	\\
$\Gamma_{f_0(500)}$ [\mevcc] &\pzz337 $\pm$ 67\pzz\pz\\
$m_{f_0(1370)}$ [\mevcc] &\pz1475 $\pm$ \pz6\pzz\pz		\\
$\Gamma_{f_0(1370)}$ [\mevcc] &\pzz113 $\pm$ 11\pzz\pz	\\	
$m_{f_0(980)}$ [\mevcc] &\pzz945 $\pm$ \pz2\pzz\pz\\
$g_{\pi\pi}$ [1/\mevcc] &\pzz199 $\pm$ 30\pzz\pz  \\
$R\frac{g_{KK}}{g_{\pi\pi}}$&\pzz3.45 $\pm$ 0.13\pzz \\
\bottomrule
\end{tabular} 
\end{center}
\end{table}

\subsection{Phase-space density}
The four-body phase-space density for the decay \decay{\Bd}{(\pipic)(\kpic)}
is parameterised by
\begin{equation}
\label{eq:phsp}
\Phi(m_{\pipin},m_{\kpin})\propto q(m_{\pipin})q(m_{\kpin})q(M_{\Bd}),
\end{equation}
\noindent being $q(m_{ij})$ the relative momentum of the final-state particles in their parent rest frame,
\begin{equation}
\label{eq:q}
q(m_{ij}) = \frac{\sqrt{(m^2_{ij}-(m_i + m_j)^2) (m^2_{ij}-(m_i - m_j)^2)}}{2m_{ij}}.
\nonumber
\end{equation}
\subsection{Relativistic Breit--Wigner}
This shape is given, as a function of the two-body invariant mass, $m$, and the relative angular momentum between, $L$, among the two decay products by
\begin{eqnarray}
\label{eq:BW}
BW(m,L) = \frac{m_0\Gamma_0}{m_0^2-m^2 -i m_0 \Gamma_L(m)},
\end{eqnarray}
\noindent where
\begin{eqnarray}
\label{eq:GammaBW}
\Gamma_L(m)&=&\Gamma_{0} \left(\frac{m_{0}}{m}\right) B_L(q,q_0,d_R)^2\left( \frac{q}{q_{0}} \right)^{2 L + 1}, \nonumber
\end{eqnarray}

\noindent being $d_R$ the radius of the resonance, and $m_0$ and $\Gamma_0$ its Breit--Wigner mass and natural width, as shown in \tab{systprops}.

\subsection{The Gounaris--Sakurai function}
 This parameterisation takes the form
\begin{eqnarray}
GS(m) \propto
\frac{1}{m_{\rho^0}^2-m^2+\Gamma_{\rho^0}\frac{m_{\rho^0}^2}{k_{\rho^0}^3}[k^2(h-h_{\rho^0})-(m^2-m_{\rho^0}^2)k_{\rho^0}^2
h'_{\rho^0}] - i m_{\rho^0}\Gamma(m)},
\end{eqnarray}
\noindent with
\[
k \equiv k(m) = (m^2/4-m_{\pi}^2)^{1/2}, \quad 
h \equiv h(m) = \frac{2}{\pi} \frac{k}{ m} \log \left( \frac{m+2 k}{2m_{\pi}} \right), 
\quad h'(m) \equiv \frac{d h(m)}{d m^2},
\]
\[
k_{\rho^0}\equiv k(m_{\rho^0}),
\quad h_{\rho^0}\equiv h(m_{\rho^0}), 
\quad \Gamma(m)\equiv\Gamma_1(m), \nonumber
\]
\noindent where $\Gamma_{\rho^0}$ is the $\rho^0$ natural width, $m_{\rho^0}$ is the $\rho^0$ Breit--Wigner mass and $d_R$ the effective radius (range parameter) of this meson, shown in \tab{systprops}.

\subsection{The Flatté parameterisation} 
This shape is described by 
\begin{eqnarray}
F(m)&=&\frac{m_0(g_{\pi\pi}\rho_{\pi\pi}(m_0)+g_{KK}\rho_{KK}(m_0))}{m_0^2-m^2-im_0(g_{\pi\pi}\rho_{\pi\pi}(m)+g_{KK}\rho_{KK}(m))}, \\
\rho_{XX}(m)&=&
  \left\lbrace
  \begin{array}{r}
  \sqrt{1-4\frac{m_X^2}{m^2}} \text{ for } m>2m_X, \\
  i\sqrt{4\frac{m_X^2}{m^2}-1} \text{ for } m\leq 2m_X,
  \end{array}
  \right. \nonumber
\end{eqnarray}
\noindent where $m_X=m_K,m_{\pi}$, accordingly. The resonance mass is represented by $m_0$ and $g_{\pi\pi}$ ($g_{KK}$)
stand for the strength of the coupling to the \decay{f_0(980)}{\pipic} (\decay{f_0(980)}{\Kp\Km}) decay channels. Their values are given in \tab{systprops}.

\newpage

\cleardoublepage
\addcontentsline{toc}{section}{References}
\setboolean{inbibliography}{true}
\bibliographystyle{LHCb}
\bibliography{main,LHCb-PAPER,LHCb-CONF,LHCb-DP,LHCb-TDR} 

\cleardoublepage
\centerline{\large\bf LHCb collaboration}
\begin{flushleft}
\small
R.~Aaij$^{28}$,
C.~Abell{\'a}n~Beteta$^{46}$,
B.~Adeva$^{43}$,
M.~Adinolfi$^{50}$,
C.A.~Aidala$^{78}$,
Z.~Ajaltouni$^{6}$,
S.~Akar$^{61}$,
P.~Albicocco$^{19}$,
J.~Albrecht$^{11}$,
F.~Alessio$^{44}$
% ...

{\footnotesize \it
% ...
$ ^{6}$Clermont Universit{\'e}, Universit{\'e} Blaise Pascal, CNRS/IN2P3, LPC, Clermont-Ferrand, France\\
% ...
$ ^{11}$Fakult{\"a}t Physik, Technische Universit{\"a}t Dortmund, Dortmund, Germany\\
% ...
$ ^{19}$INFN Laboratori Nazionali di Frascati, Frascati, Italy\\
% ...
$ ^{28}$Nikhef National Institute for Subatomic Physics, Amsterdam, Netherlands\\
% ...
$ ^{43}$Instituto Galego de F{\'\i}sica de Altas Enerx{\'\i}as (IGFAE), Universidade de Santiago de Compostela, Santiago de Compostela, Spain\\
$ ^{44}$European Organization for Nuclear Research (CERN), Geneva, Switzerland\\
% ...
$ ^{46}$Physik-Institut, Universit{\"a}t Z{\"u}rich, Z{\"u}rich, Switzerland\\
% ...
$ ^{50}$H.H. Wills Physics Laboratory, University of Bristol, Bristol, United Kingdom\\
% ...
$ ^{61}$University of Cincinnati, Cincinnati, OH, United States\\
% ...
$ ^{63}$Syracuse University, Syracuse, NY, United States\\
% ...
$ ^{78}$University of Michigan, Ann Arbor, United States, associated to $^{63}$\\
% ...
}
\end{flushleft}

\end{document}